\documentclass[11pt,
  aps,
  pra,
  superscriptaddress,
  showpacs,
  tightenlines,
  longbibliography,
  nofootinbib
]{revtex4-2}

\usepackage[margin=0.75in]{geometry} 

\usepackage{amsmath,amsthm,amssymb,amsfonts,mathtools,float,graphics,epsfig,epstopdf,color,verbatim,tabularx,bm,multirow,appendix}
\usepackage{booktabs,tabularx,array,xcolor}
\usepackage{tabularx}
\usepackage[utf8]{inputenc}
\usepackage[T1]{fontenc}
\usepackage{xcolor}
\usepackage{booktabs}
\usepackage{dsfont}
\usepackage{textcomp}
\usepackage{yfonts}
\usepackage{footnote}
\usepackage{bm}
\usepackage{subfigure}
\usepackage{amsmath}
\usepackage{mathrsfs}
\usepackage{MnSymbol}

\usepackage{mathtools}
\usepackage{graphicx}
\usepackage{verbatim}
\usepackage[colorlinks=true, citecolor=blue, linkcolor=blue, urlcolor=blue]{hyperref}
\usepackage{multirow}
\usepackage{braket}
\usepackage[normalem]{ulem}
\usepackage{tikz}
\usetikzlibrary{calc}
\usetikzlibrary{shapes.multipart}
\usepackage{orcidlink}
\usepackage{xr-hyper} 
\usepackage[percent]{overpic} 
\newcommand{\ketbra}[2]{|#1\rangle\langle#2|}

\newtheorem*{importantstatement}{Key} 

\definecolor{stab}{HTML}{6B3FA0}
\definecolor{ppt}{HTML}{168B72}
\definecolor{ink}{HTML}{18212B}
\definecolor{muted}{HTML}{66717F}
\definecolor{pale}{HTML}{F1F3F5}

\usepackage{comment}
\DeclareMathOperator{\Tr}{Tr}

\newcommand{\comments}[1]{}

\renewcommand{\cal}[1]{\mathcal{#1}}

\newcommand{\STAB}{\mathrm{STAB}}
\newcommand{\id}{\mathbf{1}}
\newcommand{\ev}[1]{\langle #1\rangle} 
\newcommand{\Wit}{\mathcal{W}}

\newcommand{\Var}{\operatorname{Var}}

\newcommand{\Nfr}{\mathcal N_{\rm fr}}
\newcommand{\nfr}{n_{\rm fr}}
\newcommand{\K}{\mathcal K}
\newcommand{\Ceff}{C_{\rm eff}}
\newcommand{\dd}{\mathrm d}
\newcommand{\smref}[1]{Sec.~\ref{#1}}


\graphicspath{{./Figures/}}
\usepackage{helvet}                      

\newcommand{\panellabel}[1]{{\fontfamily{phv}\fontseries{b}\selectfont
                             \fontsize{10}{10}\selectfont #1}}
\newsavebox{\mainpanelbox}

\usetikzlibrary{calc}
\IfFileExists{DejaVuSans.sty}{\usepackage{DejaVuSans}\newcommand{\rowfont}{\fontfamily{DejaVuSans-TLF}\selectfont}}
                             {\usepackage{helvet}\newcommand{\rowfont}{\fontfamily{phv}\selectfont}}
\newcommand{\panelletter}[1]{{\rowfont\bfseries\fontsize{8.3}{8.3}\selectfont #1}}   
\newsavebox{\boxD}\newsavebox{\boxE}\newsavebox{\boxF}

\newsavebox{\panbA}\newsavebox{\panbB}\newsavebox{\panbC}\newsavebox{\panbD}
\newcommand{\figletterpt}{10}\newcommand{\figtitlept}{8.6}   
\newcommand{\figletter}[1]{{\rowfont\bfseries\fontsize{\figletterpt}{\figletterpt}\selectfont #1}}
\newcommand{\figtitle}[1]{{\rowfont\bfseries\fontsize{\figtitlept}{\figtitlept}\selectfont #1}}

\usetikzlibrary{shapes.geometric}                 
\definecolor{hcorange}{HTML}{FF7F0E}              

\DeclareRobustCommand{\hcdiamond}{%
  \tikz[baseline=-0.5ex]\fill[hcorange] (0,0.33em)--(0.33em,0)--(0,-0.33em)--(-0.33em,0)--cycle;}

\DeclareRobustCommand{\onsagerstar}{%
  \tikz[baseline=-0.5ex]\node[star,star points=5,star point ratio=2.618,
        draw=black,fill=yellow!85!orange,line width=0.4pt,inner sep=0pt,minimum size=0.9em]{};}

\begin{document} 
\title{Fate of Quantum Magic} 
\begin{abstract}
   Magic, a quantum resource distinct from entanglement, 
   measures departure from states stabilized by products of Pauli operators.
   Such stabilizer states can be maximally entangled yet admit efficient classical description:  quantum  advantage requires magic.  
    We develop a powerful framework to evaluate the degree of magic in generic many-body states. We use it to study the fate of magic under thermal and dynamical evolution. In the transverse field Ising model, we find a thermal sudden death of magic that saturates with system size, and occurs after the death of entanglement. 
    We reach
similar conclusions in generic spin chains, and in the Kitaev quantum spin liquid on
the honeycomb lattice, where the threshold shows an intricate dependence on subregion
shape. 
    We then study how magic equilibrates after a quantum quench. 
    In quenches 
    of generic spin chains, the magic of a subregion typically dies before entanglement, contrary to thermal trajectories.
An explanation is given in terms of the geometry of non-magical states. As a third type of trajectory, we examine the fate of magic with separation with an appropriate notion of non-local magic and establish a sudden death.    
Finally, we show how to experimentally witness a volume law of frame-robust magic in quantum materials by using neutron scattering.
\end{abstract}

\author{Liuke Lyu}
\affiliation{D\'epartement de Physique, Universit\'e de Montr\'eal, Montr\'eal, QC H3C 3J7, Canada}
\affiliation{Department of Physics, University of Toronto, Toronto, Ontario M5S 1A7, Canada}

\author{Erik S. S{\o}rensen\,\orcidlink{0000-0002-5956-1190}}
\email{sorensen@mcmaster.ca}
\affiliation{Department of Physics and Astronomy, McMaster University, Hamilton, Ontario L8S 4M1, Canada}

\author{William Witczak-Krempa}
\email{w.witczak-krempa@umontreal.ca}
\affiliation{D\'epartement de Physique, Universit\'e de Montr\'eal, Montr\'eal, QC H3C 3J7, Canada}
\affiliation{
 Institut Courtois, Universit\'e de Montr\'eal, Montr\'eal (Qu\'ebec), H2V 0B3, Canada
}
\affiliation{
 Centre de Recherches Math\'ematiques, Universit\'e de Montr\'eal, Montr\'eal, QC, Canada, HC3 3J7
}
\date{\today}

\maketitle


\section{Introduction}\label{sec:intro}
Entanglement is a non-local quantum correlation that permeates many-body systems~\cite{Amico2008Review}, and 
serves as a resource for tasks that are classically limited or forbidden. 
As a subsystem interacts with its environment and becomes entangled with it, its internal entanglement can decrease and even vanish at a finite time, temperature or separation; see Ref.~\cite{FoE} and reference therein for a general analysis of the fate of multiparty entanglement. 
While entanglement quantifies nonlocal quantum correlations, a distinct resource called the quantum magic quantifies departure from classically simulable states.
Here we study the fate of quantum magic, i.e.\/ non-stabilizerness~\cite{Bravyi2005,Veitch2012,Howard2014}.
Stabilizer states admit compact descriptions in terms of commuting Pauli operators~\cite{Gottesman1998}, and Clifford evolution preserves this structure, allowing efficient classical simulation~\cite{Gottesman1998, Aaronson2004}. Suitable nonstabilizer states can, in turn, make Clifford computation universal~\cite{Bravyi2005}.
A mixed state is magic if it cannot be reproduced by a classical mixture over stabilizer
descriptions~\cite{Veitch2014,Howard2017}. 
Magic and entanglement are clearly distinct resources: a Bell pair is both maximally entangled and a stabilizer state, while a product of single-qubit magic states is separable.

In many-body physics magic was established as a generic, extensive resource~\cite{Liu2022}, quantified in ground states of spin
chains~\cite{Sarkar2020,White2021,Oliviero2022}, and linked to the symmetry-protected sign problem and the classification of phases~\cite{Ellison2021}.
Commonly used measures include the robustness of
magic~\cite{Howard2017,Heinrich2019,Hamaguchi2024} and the stabilizer R\'enyi
entropy~\cite{Leone2022}. The latter is cheaper to evaluate but has a crucial limitation: it is a magic monotone only for pure states (when R\'enyi index $\alpha\ge2$)~\cite{Haug2023,Leone2024}.
Scalable estimators exist for both~\cite{Tarabunga2023,Lami2023,Timsina2025}.
Magic can also be non-local, in the sense that it is a property of the system as a whole that cannot be removed by acting on its parts separately. This idea has been formalized in inequivalent ways. Long-range magic is magic that survives every shallow local circuit; it is connected to phases of matter~\cite{Korbany2025,Wei2025}. For a bipartition, non-local magic is what remains after minimizing over local unitaries on each party~\cite{Cao2025, QianWang2025}; it is bounded by properties of the entanglement spectrum and controls gravitational backreaction in holographic CFTs~\cite{Cao2025}. Magic can also be entirely non-local in the sense that every marginal is non-magical while the joint state is magical, which can occur even for separable states~\cite{Wei2026}. Thus magic and entanglement are not ordered in any simple way.
Most of these studies concern pure states at zero temperature, whereas for mixed states
even deciding membership in the stabilizer polytope $\STAB$ is computationally
hard~\cite{Leone2026}. In odd Hilbert space dimension, like spin-1, an efficient sufficient but not necessary criterion exists~\cite{Gross_2006}.
Here, we tackle the challenging regime of mixed qubit states, quantifying magic by the Bures distance to $\STAB$, a monotone that vanishes only for non-magical states. Our approach operates directly on full-rank mixed states of many qubits and returns
not merely the presence or absence of magic but a quantitative value.

We focus on the study of quantum magic during thermal and dynamical evolution in local Hamiltonians. This is done on several fronts. First, we develop SOFA, a scalable algorithm that accurately evaluates a magic monotone, the Bures distance to $\STAB$ derived from the  fidelity~\cite{Bravyi2019,Warmuz2024}, for generic mixed states. We reach $12$-qubit mixed states in challenging regimes (near the death of magic), but going beyond becomes increasingly challenging. At the same time, obtaining large reduced density matrices is often not practical. These constraints motivate the quantum microscopy (or entanglement microscopy) approach~\cite{Wang2025entanglement}: extract quantum correlations of sufficiently small subregions.  For the 1d transverse field Ising model (TFIM) we find sharp thermal thresholds, allowing us to identify the finite temperature at which magic dies for the entire chain in the thermodynamic limit. This thermal sudden death in the TFIM was not resolved by a previous study~\cite{Timsina2025}, presumably because of algorithmic limitations in the
challenging near-threshold regime. The entanglement death temperature is also
identified, and is strictly smaller than that of magic. We then study the quantum spin liquid in the Kitaev honeycomb model~\cite{Kitaev2006} at finite temperature, and again find entanglement-first death, with an intricate dependence on subregion shape. A further analysis of generic spin chains reaches the same conclusion. We thus conclude that in typical thermal evolution entanglement dies before magic. In quantum quenches we observe the opposite ordering: during dynamical evolution magic typically dies first. We argue from the geometry of stabilizer states that the death ordering found for quenches is the typical one, and explain the special nature of thermal trajectories. Next, we examine the fate of magic along a different kind of trajectory, spatial separation, by introducing an appropriate notion of non-local magic, dubbed ``Bell producibility,'' inspired by genuine network multipartite entanglement~\cite{navascues2020genuine}, and establish a sudden death with separation. Finally, we show that a \emph{volume law of frame-robust quantum magic} arises whenever a local signature persists throughout homogeneous matter, even at finite temperature or with state preparation errors; the resource it certifies is the stabilizer nullity~\cite{Beverland2020}. 
This volume law can be witnessed with the same neutron-scattering observables used for entanglement witnesses~\cite{Hauke2016,Scheie2021,Scheie2025Tutorial}, and we provide predictions for the 1d and 2d TFIM. A strong magic witness for the Kitaev honeycomb spin liquid is also discussed.

\section{The geometry and death of magic }\label{sec:geometry}

Clifford quantum circuits map strings of Pauli operators to other such strings under conjugation~\cite{Gottesman1998,Aaronson2004}. Starting from spins polarized along specified Pauli axes (the frame), they generate the class of stabilizer states~\cite{Gottesman1996,Gottesman1997}. Magic denotes departure from the convex set of their statistical mixtures, relative to that Pauli frame.
The set of states without magic, STAB, has nonzero volume in the space of all states,
just like the set of unentangled states, SEP~\cite{Zyczkowski}.
We recall that STAB is defined for $n$ qubits as the convex hull of pure stabilizer states $U|0\rangle^{\otimes n}$, where $U$ is a Clifford gate. SEP is defined analogously as the convex hull of $n$-qubit product states $|\psi_1\rangle\otimes\cdots\otimes |\psi_n\rangle$.
Both sets overlap (they contain the maximally mixed state) but are clearly distinct: Bell or GHZ states belong to STAB but are entangled.

\begin{figure}
    \centering
    \includegraphics[width=.5\linewidth]{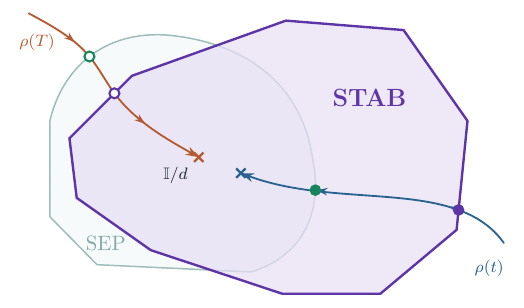}
    \caption{Set of stabilizer states without magic (STAB), and the set without entanglement (SEP). Typical late-stage thermal ($T$) and temporal ($t$) evolutions are shown in red and blue, respectively.
    STAB-first entry is the typical case. The infinite-temperature maximally mixed state is $\mathbb I/d$. }
    \label{fig:geometry}
\end{figure}

We shall be concerned with the fate of magic under some evolution $s$, of the reduced density matrix (RDM) of a subregion, $\rho(s)$, where $s$ could be temperature, time, or separation. If the evolution ends inside STAB, then there will be a definitive sudden death of magic, as illustrated in Fig.~\ref{fig:geometry}. 
This mirrors the fate of entanglement~\cite{FoE}
and raises the question of which quantity typically dies
first, magic or entanglement.
Below, we shall provide sharp thresholds for both resources.

We begin by recalling a key fact about the STAB polytope: most of its vertices are entangled
because the total number of stabilizer states grows like \(2^{n^2/2+O(n)}\), whereas the number of product stabilizers  grows as \(6^n\) ($99.99\%$ of 6-qubit vertices are entangled)~\cite{Aaronson2004}. Since mixtures of entangled states can be separable, one needs to go beyond the vertices. We sample the boundary isotropically by generating Haar-induced mixed states and projecting them to the boundary; see Methods. Working with 6 qubit states, 99.5\% of sampled boundary points are entangled. Therefore, most of the boundary of STAB is entangled, which in turn implies:
\begin{importantstatement}
A typical trajectory $\rho(s)$ will enter STAB before SEP: magic dies first, then entanglement. 
\end{importantstatement}

In order to quantitatively determine the fate of magic, our main tool will be the
Bures distance from a state $\rho$ to STAB, $d_B$, determined by the stabilizer
fidelity introduced for pure states in Ref.~\cite{Bravyi2019}. Here we work with its mixed-state form, the Uhlmann fidelity maximized over $\STAB$, 
which places $d_B$ among the faithful mixed-state magic monotones, alongside those built from quantum relative entropies~\cite{Veitch2014,Rubboli2024} and from the
boundary structure of the polytope~\cite{Warmuz2024}.
In contrast, the monotonicity of the stabilizer
R\'enyi entropy~\cite{Leone2022} 
only holds on pure states~\cite{Haug2023,Leone2024}.
$d_B$ is defined by minimizing distance, or equivalently by maximizing the mixed state (Uhlmann) fidelity $F$:
\begin{align} 
   d_B^2(\rho)=\min_{\sigma\in{\rm STAB}} 2(1-\sqrt F)\,, \qquad F=\left(\Tr\sqrt{\sqrt\sigma \rho \sqrt\sigma} \right)^2
\end{align}
Crucially, $d_B$ is monotone under stabilizer-preserving CPTP maps, since the Uhlmann fidelity is non-decreasing under any CPTP map~\cite{nielsen2010quantum}, and a stabilizer-preserving map by definition sends $\STAB$ into itself.

{\bf Algorithm}---We developed a scalable approach for magic detection, Stabilizer-Orbit Frank-Wolfe algorithm (SOFA).
SOFA minimizes the squared Bures distance $d_B^2$ to the
 stabilizer polytope using a GPU-accelerated, fully corrective Frank--Wolfe~\cite{Frank1956} method. Its approximate linear minimization oracle searches Clifford orbits in parallel for stabilizer vertices predicted to produce the largest decrease in $d_B^2$; the weights of all active vertices are then jointly re-optimized over the probability simplex using a safeguarded
   projected quasi-Newton method. This approach enables calculations for systems of up to 12 qubits at temperatures approaching the sudden-death threshold, where the
  state not only remains full rank, but numbers of vertices needed becomes substantial.
SOFA yields a rigorous upper bound for $d_B$. The quality of the bound can be judged by monitoring the rate of decrease of distance per iteration, the increase in oracle strength, and stability with respect to the addition of large pools of candidate vertices (as obtained from a nearby temperature, say).

\section{Fate with temperature}\label{sec:FatewT}
We begin with the fate of magic under thermal evolution. Let us consider an equilibrium Gibbs state $\rho=\exp(-\beta H)/\mathcal Z$ of a local Hamiltonian $H$; $\beta=1/T$ is the inverse temperature. 
 Since the infinite temperature state, $\rho=\mathbb{I}/d$, is maximally mixed, and $\STAB$ contains a finite-radius ball around
it~\cite{liu2026geometrytypicalityquantummagic}, a sudden death of
magic will occur at some temperature $T_c$.

\begin{figure}
    \centering
    \includegraphics[width=.90
    \linewidth]{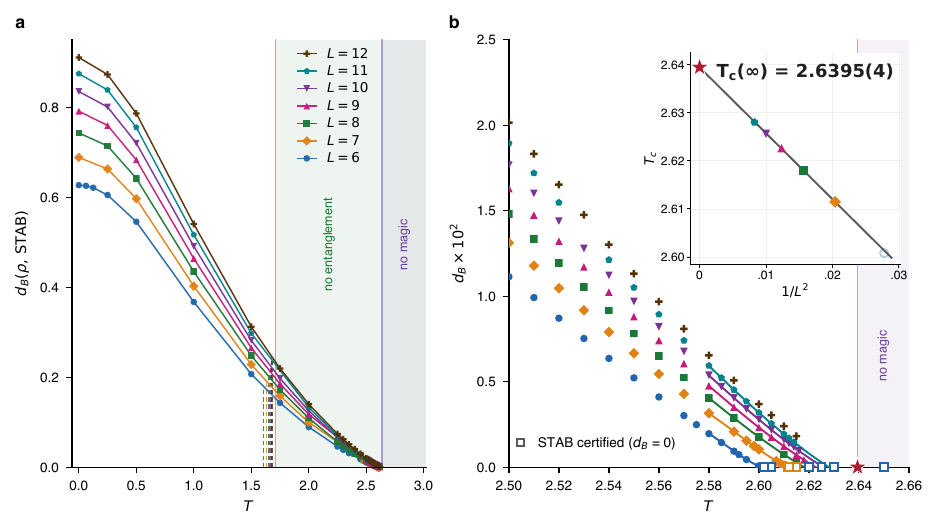}
    \caption{Finite temperature evolution of magic in subregions of $L$ sites in the infinite transverse field Ising model as measured by the Bures distances to STAB. a) Full temperature range.
    The vertical dashed lines denote the full-PPT temperatures for each $L$. b) Near-death scaling with cubic fits $d_B=a x+b x^2+c x^3$, where $x=T_c-T$. $L=12$ is not fitted due to looser convergence near threshold.  Inset: extrapolation of $T_c$ to the thermodynamic limit. }
    \label{fig:ising}
\end{figure}

{\bf Quantum Ising}---Our central example is the 1d TFIM:
\begin{align}
    H_{\rm TFIM} = \sum_i (-J\sigma_i^x \sigma_{i+1}^x - h \sigma_i^z) 
\end{align} 
This model is exactly  solvable~\cite{Pfeuty1970} and has a quantum
critical point at $h_c=J$ separating a low-$h$ ferromagnet ($J>0$) from a high-$h$
paramagnet. Its ground-state magic was among the first many-body cases analyzed in resource-theoretic terms~\cite{Sarkar2020, Oliviero2022}.
We set $J=1$ and work with the infinite chain at temperature $T$; we consider the reduced density matrix for $L$ contiguous sites as our state $\rho(T)$ (see Methods). Figure~\ref{fig:ising} shows the $T$-dependence of $d_B$ obtained with SOFA for different block sizes. The striking feature is the rapid clustering of the curves near the sudden death temperature of magic. For $L=6,7$, we can certify STAB membership exactly as denoted by squares in panel b (see Methods and Supplemental Materials~\ref{SM:stab_membership}), which independently validates the death of $d_B$ extracted by the extrapolations. 
The restriction to small blocks is intrinsic: deciding $\STAB$ membership, even
approximately, requires super-exponential time under the exponential time
hypothesis~\cite{Leone2026}.
This explains why our tight STAB certificates stop at $L=7$.
In the inset of Fig.\ref{fig:ising}, we extrapolate $T_c(L)$ to the thermodynamic limit and obtain $T_c(\infty)=2.6395$: the temperature at which the infinite chain becomes entirely devoid of magic, a threshold not resolved by the quantum Monte Carlo tomography of the robustness of magic in the same model~\cite{Timsina2025}. 
Up to $L=7$, where an exhaustive search over stabilizer states is feasible, we also construct explicit witness which certifies magic just below the threshold (see Methods). Since tracing out sites cannot create magic, $T_c(L)$ is non-decreasing in $L$, we can establish a rigorous lower bound $T_c(\infty)\geq T_c(L=7)>2.61$.
Further, we have the following operational implication: a device using thermal TFIM states as its only non-stabilizer resource must operate below $T_c$ to supply magic. 

We can compare the observed evolution to the fate of entanglement~\cite{FoE}, which we study by finding the temperature at which a block acquires a positive partial transpose~\cite{Peres1996,Horodecki1996PPT}  (PPT) for all bipartitions. Remarkably, for $L\leq 7$ we find that the PPT matches the full death of entanglement by establishing rigorous SEP membership for temperatures at most $10^{-3}$ above $T_{\rm PPT}$. Coherent with its NP-hard computational class~\cite{gharibian2008strong}, the certification becomes algorithmically more challenging with an increasing number of spins, but we can nevertheless certify temperatures as low as $1\%$ above PPT for $L=9$, and at $L=12$ $T=1.8293$ as exactly separable. Building on our exact upper bounds, we conjecture that $T_{\rm PPT}=T_{\rm SEP}$ for all $L$, and by extrapolating we obtain, $T_{\rm PPT}(\infty) = 1.7093$ which we claim equals $T_{\rm SEP}(\infty)$, shown in green in Fig.~\ref{fig:ising}a. This is in agreement with the proof of a bounded entanglement temperature for local Hamiltonians by Bakshi \emph{et al.}~\cite{Bakshi2024}, which actually establishes a more restrictive property than death of entanglement or magic:  above some temperature the state is both in SEP and STAB, a mixture of single-site product \emph{stabilizer} states.
The bound of Bakshi \emph{et al.}~\cite{Bakshi2024} is then also a bound on magic yielding, $T > 800\,J$ for the TFIM at $|h|\le J$, a bound that is far above the threshold $T\simeq 2.6\,J$ established here.
Contrary to the typical expectation, we thus find that magic dies well after entanglement, an ordering that has only recently been questioned~\cite{Cao2026}.
Before moving to other models, we need to assess robustness of our conclusion with respect to the choice of frame, beyond the canonical one. For $L=6$, we searched over frames, and found that $T_c(6)$ could be lowered from $2.60$ down to $2.4$. By monotonicity, this shows that $T_c(\infty)\geq 2.4$, which remains substantially above the entanglement death temperature.

\begin{figure}
\centering
\begin{tikzpicture}[inner sep=0pt]
  \pgfmathsetlengthmacro{\gapx}{-3mm}                 
  \pgfmathsetlengthmacro{\wa}{0.65\linewidth}         
  \pgfmathsetlengthmacro{\wc}{\linewidth-\wa-\gapx}   
  \node[anchor=north west] (a) at (0,0)
    {\includegraphics[width=\wa]{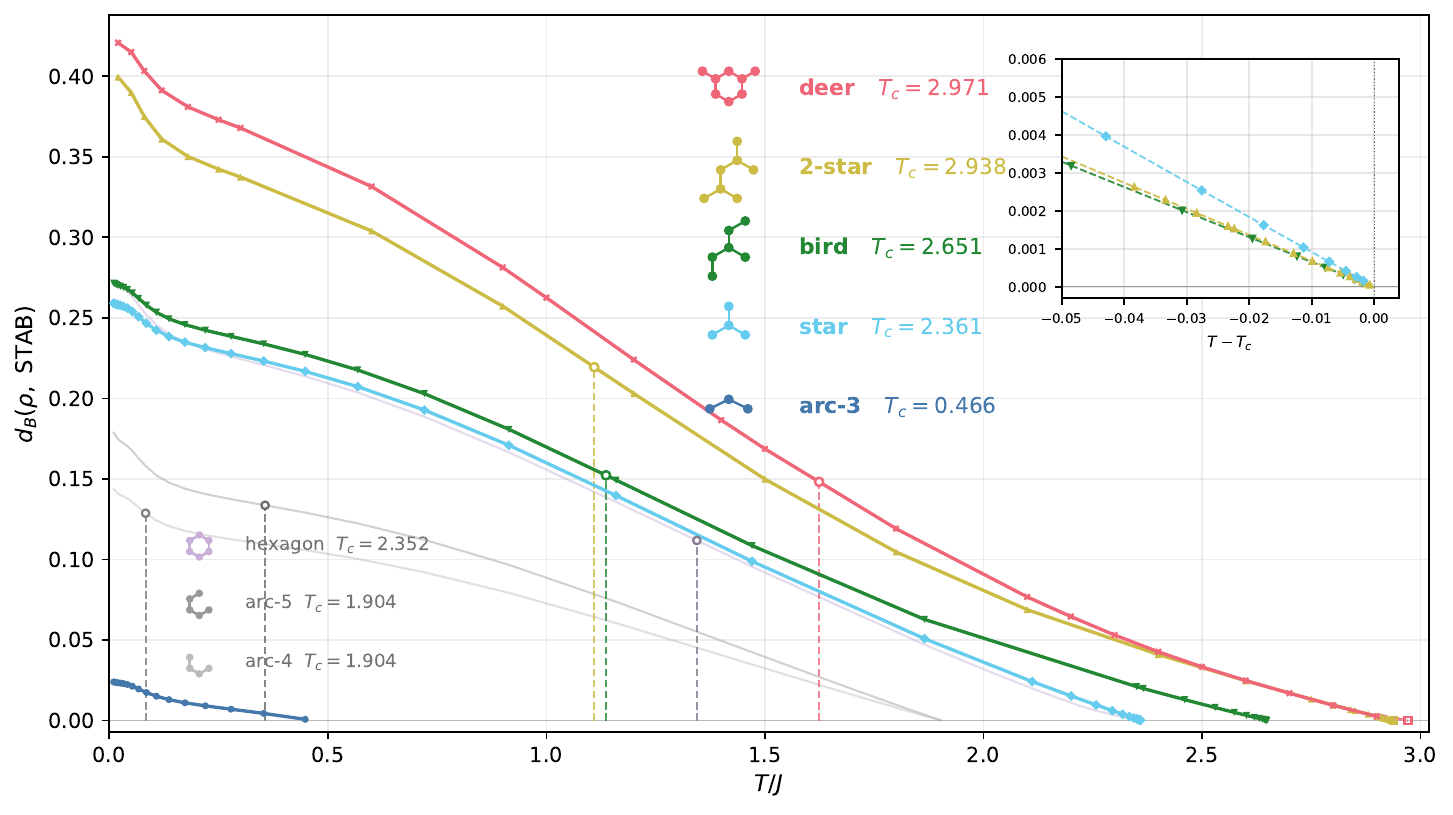}};
  \path let \p1=(a.north west), \p2=(a.south east) in
    \pgfextra{\pgfmathsetlengthmacro{\ha}{\y1-\y2}\xdef\ha{\ha}};
  \node[anchor=south west] (c) at ($(a.south east)+(\gapx,0)$)
    {\includegraphics[width=\wc]{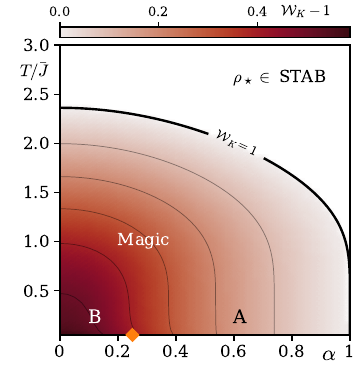}};
  \node[anchor=north west] at ($(a.north west)+(3mm,-1.5mm)$) {\panellabel{a}};
  \node[anchor=north east] at ($(a.north west)+(0.9465*\wa,-0.0726*\ha)+(-0.6mm,-0.6mm)$) {\panellabel{b}};
  \node[anchor=east] at ($(c.north west)+(8.4mm,-2.5mm)$) {\panellabel{c}};
\end{tikzpicture}
\caption{\panellabel{a} Thermal evolution of magic in the isotropic Kitaev honeycomb model for various subregions of a 32-site torus, up to 8 sites. The coloured subregions are the ones that have the most resilient magic against temperature for a given number of spins. \panellabel{b} Inset: near-death scaling of magic.
\panellabel{c} The magic witness ${\cal W}_K$ of the four-site star over the anisotropy--$T$ plane going from the B-phase (gapless) to the A-phase (gapped), $J_z=1+2\alpha$, $J_x=J_y=1-\alpha$. $\bar J$ is the average coupling. \hcdiamond{}: $\alpha_c=1/4$ at $T=0$.
From the Gibbs state of a
24-site torus. At $\alpha=0$ the critical $T_c$ for ${\cal W}_K$ is $T_c=2.361$ in agreement with \panellabel{a}.}
\label{fig:Kitaev}
\end{figure}

{\bf Kitaev spin liquid}---We consider the ferromagnetic Kitaev model on the honeycomb lattice~\cite{Kitaev2006}
$H=-\sum_{\ev{ij}_a}J_a\,\sigma^a_i\sigma^a_j$.
We first study the isotropic point where the groundstate is a spin liquid with emergent gapless Majorana fermions. We exploit the exact solution to get thermal density matrices on the highly symmetric 32-site torus (see Methods). 
The global ground-state magic of this model has only been studied via the stabilizer R\'enyi entropy~\cite{Lamma2026}.
Figure~\ref{fig:Kitaev} shows the thermal evolution of $d_B$ for various subregions: we again observe the sudden death of magic. 
The coloured subregions are the ones with the maximal $T_c$ for a given $L$, while the gray ones have less robust magic. For instance, the magic of the hexagonal plaquette is less robust than for the ``bird'' subregion at $L=6$, but its entanglement (vertical dashed lines) is more robust than the bird's. This last statement is coherent with the discovery of multiparty entanglement being predominantly encoded in closed loops in spin liquids~\cite{lyu2025loop}. Here we see that magic possesses a different structure: it resides in less compact subregions. 
In all cases, we find the death of entanglement (vertical dashed lines) occurs before the death of magic.

{\bf Generic spin models}---The Kitaev and Ising models have a rather simple structure, and are exactly solvable through fermionization, although the latter property does not necessarily lead to peculiar entanglement properties for subregions when compared to other Hamiltonians. We now investigate magic in more generic models, focusing on translation-invariant chains:
\begin{align} \label{eq:H_rnd}
    H=\sum_{i}\left( J_1^{ab}\!\sigma_i^a \sigma_{i+1}^b + J_2^{ab}\! \sigma_i^a \sigma_{i+2}^b + K^{abc}\! \sigma_i^a \sigma_{i+1}^b \sigma_{i+2}^c + \bm h \cdot\bm{\sigma}_i \right)
\end{align}
The two- and three-spin couplings $J_I, K$, and Zeeman field $\bm h$ are chosen randomly for each realisation using a Gaussian distribution with mean 0, and unit standard deviation. 
We rescale the parameters such that $||J_1||_F=1, ||J_2||_F=0.5$, $|\bm h|=0.7$ and $||K||_F=0.6$; we have used the Frobenius (Euclidean norm).
Note that the full exchange and three-spin coupling tensors are retained, so that generic realisations break spin parity, spatial reflection, and time-reversal symmetry while preserving translation invariance.
We study $24$-site open chains; the subregion is the central adjacent 6-site block for which we obtain the Gibbs reduced density
(see Methods).
Out of 50 random Hamiltonians, 42 instances show an entanglement-first death, Table~\ref{tab:resource-death-order}. Full STAB and SEP certification were performed, in addition to checking the simpler PPT criterion. We are thus led to the conclusion:
\begin{importantstatement}
Entanglement typically dies before magic as temperature increases in equilibrium Gibbs states of local Hamiltonians. 
\end{importantstatement}
This stands in contrast to typical trajectories because Gibbs states of local Hamiltonian possess low entanglement (cf.\/ the boundary law ~\cite{Wolf2008,Eisert2010AreaLaw}), and so they enter STAB through its rarer unentangled patches.

\begin{table}
\centering
\small
\begin{tabular}{lccccl}
\hline
Trajectory & $n$ & Resolved & STAB-first & SEP-first & Dominant order \\
\hline

Thermal
& 50 & 50 & 8 & {\bf 42}
& SEP-first,\;\;\, $42/50=84\%$ \\

Quenches (same)
& 50 & 40 & {\bf 36} & 4
& STAB-first, $36/40=90\%$ \\

Quenches (different)
& 30 & 25 & {\bf 24} & 1
& STAB-first, $24/25=96\%$ \\

\hline
\end{tabular}
\caption{Resource-death ordering in the trajectory ensembles of thermal evolution and quenches. Same (different) means initial and final coupling tensors were rescaled to have the same (different) relative ratios. }
\label{tab:resource-death-order}
\end{table}

\section{Fate with time}\label{sec:Fatewtime}
We now turn to the more challenging case of the fate of magic under dynamical
evolution, whose growth after a quench has been followed through the stabilizer
entropies~\cite{Rattacaso2023} and studied in circuit settings~\cite{Tarabunga2025,Fux2025}. 
We shall focus on quantum quenches: an initial state $|\psi(0)\rangle$ (the groundstate of a Hamiltonian $H_i$) time-evolves for $t>0$ under $H_f$. We will track the time-dependence of subregion of adjacent $L$ sites, $\rho(t)$. Contrary to the thermal case, we have no warranty that the late-time state will be in STAB. For generic quenches, $\rho$ should thermalize at late times when the subregion is sufficiently small compared to the complement. We can now invoke the results of the previous section: sufficiently high temperature destroys magic. We are thus lead to the expectation that if the quench injects enough energy so that the effective temperature exceeds the thermal threshold, magic should die. The same argument holds for entanglement~\cite{FoE}. 

\begin{figure}
    \centering
    \includegraphics[width=0.85\linewidth]{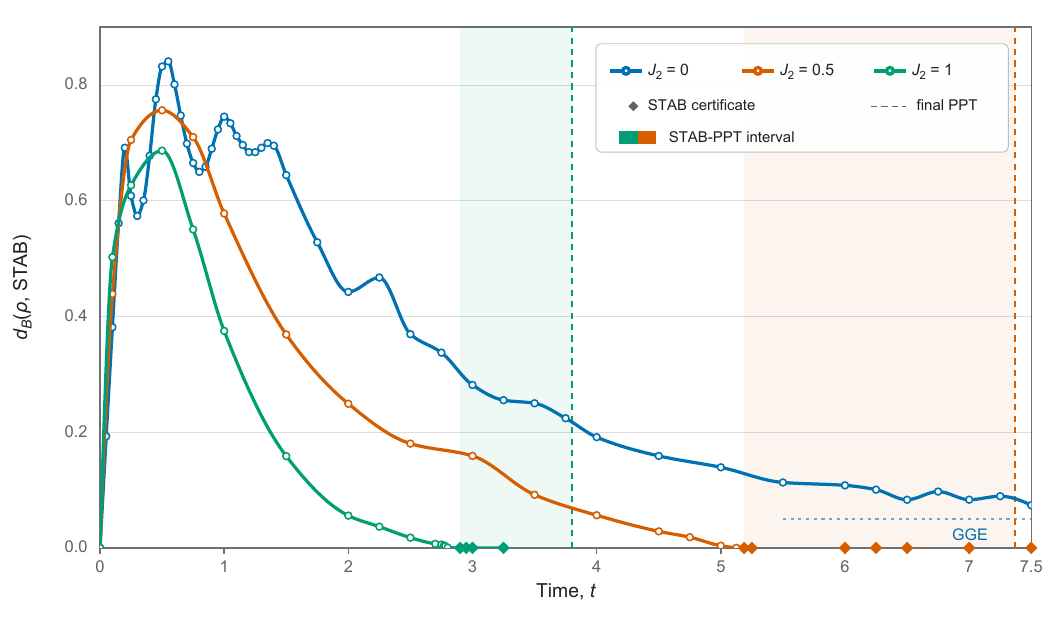}
    \caption{Dynamical fate of magic for quantum quenches in the Ising $+J_2$ spin chain; 6-site subregions are studied. The initial condition is a dimer cover of nearest-neighbour Bell states. For the integrable case $J_2=0$, we work in the thermodynamic limit, whereas we use $N=24$ for finite $J_2$.}
    \label{fig:quench}
\end{figure}

Let us first consider a quench in the 1d TFIM. The initial condition is an entangled yet non-magical dimer cover of Bell states $(|00\rangle+|11\rangle)/\sqrt2$ (a valence bond crystal of triplets). We time-evolve the state with the critical TFIM Hamiltonian by exploiting the exact Jordan-Wigner solution (see Methods). The energy-matched Gibbs reference temperature for that quench is $T_{\rm eff}=3.739$, which is well above the magic threshold temperature found above $T_c=2.6395$. The time-dependence of $d_B$ is shown in Fig.~\ref{fig:quench}: we find no sudden death of magic. We attribute this persistence of magic to the non-thermal steady state of this integrable quench: the late-time state is described by a generalized Gibbs ensemble (GGE) rather than a simple thermal state~\cite{Fagotti2013RDM}. The $d_B$ corresponding to the late-time GGE steady state is shown by a horizontal dashed line in Fig.~\ref{fig:quench}.
To get generic behavior we need to break integrability, which we do by adding a next-to-nearest neighbour coupling: $H_{\rm TFIM}+J_2\sum_i\sigma_i^x \sigma_{i+2}^x$. The results for $J_2=0.5$ and $1.0$ are shown in Fig.~\ref{fig:quench}. Crucially, $J_2>0$ recovers a sudden death of magic. Further, we find that the state becomes fully PPT strictly after the magic death: the shaded regions show the non-magical but entangled regime. 
Contrary to the thermal trajectory, this magic-first order is consistent with the geometrical expectation given above. 

In order to confirm the above, we use the same Hamiltonian family as above Eq.~(\ref{eq:H_rnd}) to perform quenches, where all couplings are chosen randomly for both $H_i$ and $H_f$. The groundstate of $H_i$ gives the initial condition, while $H_f$ is the driving Hamiltonian. We perform the same rescaling as above to preserve the relative norms in both $H_i$ and $H_f$; this will be relaxed below. We study periodic chains of $N=24$ sites via Krylov evolution. 
We find that 40 out of the 50 quenches Table~\ref{tab:resource-death-order} show a death of magic and entanglement for $t\leq 40$ (time is measured in the units where $||J_1||_F=1$); we mark the remaining 10 as `unresolved'. In a clear majority of 36 cases out of the resolved 40 we find that magic dies before entanglement. In a second 30-realisation ensemble, we further randomize the relative strengths of the coupling tensors of both initial and final Hamiltonians. Out of 25 resolved quenches, 24 show magic-first death. 
So the overall conclusion is that quenches show the expected behavior for typical trajectories, since most of the boundary of
STAB is entangled as established in Sec.~\ref{sec:geometry}
above (see also ~\cite{liu2026geometrytypicalityquantummagic}).
\begin{importantstatement}
Magic typically dies before entanglement as time increases in quantum quenches of local Hamiltonians. 
\end{importantstatement}

\section{Fate with separation}\label{sec:FatewSep}
How about the fate of magic with respect to spatial separation? To have a well-defined notion of separation we work with local Hamiltonians. 
Since magic is a global quantity, 
widely separated subregions of Gibbs states will each provide a finite contribution.
One has to define a different quantity that measures the amount of magic that cannot
be independently prepared by the parties (``non-local magic'')~\cite{Korbany2025,Wei2025,Odavic2023,Cao2025}.
Here, we introduce an even stronger notion that truly extracts the magic that cannot be produced locally.
Consider two parties, Alice and Bob. First, we allow them to share \emph{arbitrary} stabilizer states $\omega$; in particular, these resources could be of unbounded size. Second, Alice can perform arbitrary quantum channels (Completely Positive Trace Preserving maps) on her share of the received resource $\mathcal E_A$; same for Bob. So each party can inject magic locally. Finally, we allow for mixtures. We call the hull of such states \emph{STAB producible}: $\sum_k \mathcal E_A^k\otimes \mathcal E_B^k(\omega_k)$. It can be shown that in the case of qubits, one can consider the resources to be a number of shared Bell states, and ancilla qubits. So STAB producible can be renamed \emph{Bell producible}. This definition which is related to non-local magic, yet stronger, is inspired by genuine network multiparty entanglement~\cite{navascues2020genuine}. Bell producibility is clearly frame-independent.

\begin{figure}
    \centering
    \includegraphics[width=0.86\linewidth]{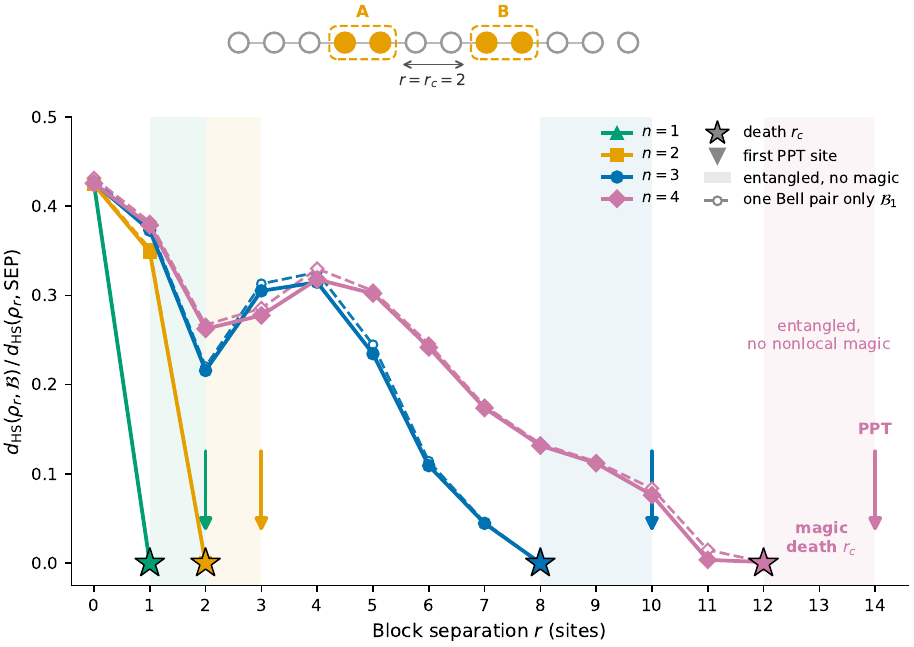}
    \caption{Death of genuine bipartite magic versus separation as measured with respect to the Bell producible hull, $d_{\rm HS}(\rho_r,\mathcal B)$; $d_{\rm HS}$ is the Hilbert-Schmidt distance. We normalise with respect to the distance to the set without bipartite entanglement, $d(\rho_r, \mbox{SEP})$. The arrows denote the death of bipartite entanglement.  }
    \label{fig:distance}
\end{figure}

As our main example, we consider two intervals of $n$ sites separated by $r$ sites in the groundstate of the TFIM, as shown in Fig.~\ref{fig:distance}. 
Instead of $d_B$ we evaluate the simpler Hilbert-Schmidt distance $d_{\rm HS}$ between the
reduced density matrix $\rho_r$ and the Bell producible set $\mathcal B$. 
We say that a state outside $\mathcal B$ has genuine bipartite magic (GBM). The key challenge is that we have to consider an increasing number of shared Bell states and ancillas in order to obtain the strongest upper bound. Figure~\ref{fig:distance} shows our upper bound for $d_{\rm HS}$ normalised by the distance to the set without entanglement (SEP) between Alice and Bob for $n=1,2,3,4$. We used up to $n_B=n$ shared Bell pairs, and up to $k=4$ ancilla qubits per party. Interestingly, the single-Bell pair plus ancillas already captures the majority of the signal at most separations.
In all cases, we find a death of GBM strictly before the death of entanglement. The latter is established via the vanishing of $d(\rho_r,{\rm SEP})$ within machine precision, which we find coincides with the separation at which PPT arises and rules out bound entanglement.
The latter was argued to occur generically beyond some separation $r_c$~\cite{FoE,Parez2024}, which then forces a death of GBM since $\mathcal B$ includes SEP. Since the Bell-producible set $\mathcal B$ strictly includes SEP, and is in fact much bigger by virtue of containing many entangled states, we have:
\begin{importantstatement}
Genuine bipartite magic (GBM) measures the amount of magic that the parties cannot prepare given arbitrary stabilizer resources and party-local quantum channels, including ones that inject magic locally. GBM has a sudden death with separation, which occurs typically before entanglement death.  
\end{importantstatement}
We can naturally generalize STAB producibility to multiple parties. We distribute an arbitrary STAB resource among $k$ parties, and then allow them to perform arbitrary party-local quantum channels. 
Allowing mixtures gives a convex set that contains all fully separable states, including those with local magic, as well as many multipartite entangled states such as GHZ states.
A state outside this set has genuine multiparty magic (GMM). 
We leave this concept for future investigation. 

\section{Experimental probes of magic}\label{sec:experiments}

 Magic has been measured on programmable quantum
processors~\cite{Oliviero2022b,Haug2023b,Niroula2024}, but none of the protocols used
are available in a quantum material.
In this section, we therefore provide criteria for detecting magic in quantum materials. We first present an experimentally-relevant witness that detects magic in the Kitaev model.
The star subregion (Fig~\ref{fig:Kitaev}a) of the honeycomb model admits such a simple yet powerful witness. Define
 $\Wit_{K}=\sum_{a=x,y,z}|c_a|,$ $ 
 c_a=\ev{\sigma^a_i\sigma^a_{j_a}}_{a\text{-bond}}$. It can be shown (see \ref{SM:Wk}) that
$
 \rho_\star\in\STAB\iff \Wit_K\le1.
$
Other, even simpler 2-site witnesses, in the spirit of recent few-measurement
magic criteria for mixed states~\cite{Haug2026,Varela2026,Tang2026,Liu2026},
are discussed in \ref{SM:frame-witness}.
At the isotropic point $\Wit_K\le 1$ is then equivalent to $|e|\le J/2$, with $e$ the energy per site and $\Wit_K$ is
then an instance of an \emph{energy witness}~\cite{Macedo2026}.
Results for $\Wit_K$ are shown in Fig~\ref{fig:Kitaev}c as a function of $T$ and anisotropy $\alpha$, matching the critical
temperature ($\Wit_K=1$) of $2.361J$ at the isotropic point from Fig~\ref{fig:Kitaev}a, showing a marked decrease in $\Wit_K$ as the transition between the A and B phases in the Kitaev model is crossed at $\alpha=1/4$, disappearing at $\alpha=1$. The considerable advantage of this witness is that it only requires knowledge of local bond energies; on the other hand, it requires knowledge of the local Hamiltonian axis (Pauli frame) which may not be available in an experimental setting. A witness independent of the frame would be of considerable interest.

Here, we establish a \emph{volume law of frame-robust quantum magic} whenever a local signature persists throughout homogeneous matter, even in the presence of a finite temperature or state preparation errors. Consider a translation-invariant spin-$1/2$ crystal with one magnetic site per primitive cell, in any dimension. Measure $m=\langle\sigma_i^z\rangle=\langle\sigma_j^z\rangle$ and $C_1=\langle\sigma_i^x\sigma_j^x\rangle$ for equivalent nearest neighbours, using orthogonal calibrated axes. Magnetometry gives $m$; the spatial Fourier component of energy-integrated, normalised magnetic neutron scattering gives $C_1$~\cite{Scheie2021,Laurell2025,Scheie2025Tutorial}. Qubit simulators obtain both by readout in two bases, without tomography. 

The spin-squeezing inequality $m^2+|C_1|\leq1$~\cite{Toth2009,sorensen} also holds, as we prove, for \emph{every mixture of independently locally rotated stabilizers}. This set includes all product and maximally entangled pure pairs in arbitrary frames. Violation excludes even explanations using different frames in each mixture component. Our local-to-global theorem gives
\begin{equation}
 m^2+|C_1|>1
 \quad\Longrightarrow\quad
 \boxed{n_{\rm fr}\geq2\bigl(m^2+|C_1|-1\bigr)>0.}
 \label{eq:experimental-volume-law}
\end{equation}
Here $n_{\rm fr}=\mathcal N_{\rm fr}/L$ is the minimum ensemble-averaged \emph{stabilizer nullity} per spin, after optimizing all pure-state decompositions and independent one-spin frames. Nullity counts the logical qubits carrying non-stabilizer information after Clifford compression~\cite{Gottesman1997,Beverland2020}. Its operational strength is direct: $\mathcal N_{\rm fr} > k$ rules out preparation whose
only non-stabilizer input is a register of at most $k$ qubits, even with Clifford
gates, stabilizer ancillas, Pauli measurements, and feed-forward.
A finite positive bound (\ref{eq:experimental-volume-law}) forces $\mathcal N_{\rm fr}$ to grow linearly with $L$ and requires extensive non-stabilizer inputs.
The proofs, along with stronger bounds and lattice prefactors are in the SM.
Crucially, resolving the remaining diagonal spin correlations strengthens the certificate without changing its frame-robust or volume-law interpretation. Replacing \(|C_{xx}|\) by \(\tfrac12[1-C_{zz}+|C_{xx}-C_{yy}|]\), which is never smaller, improves both detection and the quantitative preparation-cost bound.
The additional correlation channels require the corresponding polarization reconstruction and calibration. That can be substantially more demanding than reconstructing one scalar correlation; converting the measured cross-section into resolved spin components is already a central experimental step~\cite{Moon1969,Scheie2025Tutorial}.
\begin{figure}
\centering
\begin{tikzpicture}[inner sep=0pt, outer sep=0pt]
  \pgfmathsetlengthmacro{\gap}{0.015\linewidth}
  \sbox{\panbA}{\includegraphics{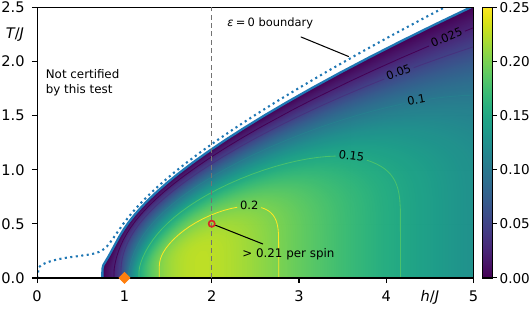}}
  \sbox{\panbB}{\includegraphics{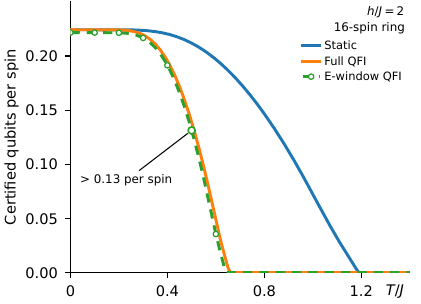}}
  \sbox{\panbC}{\includegraphics{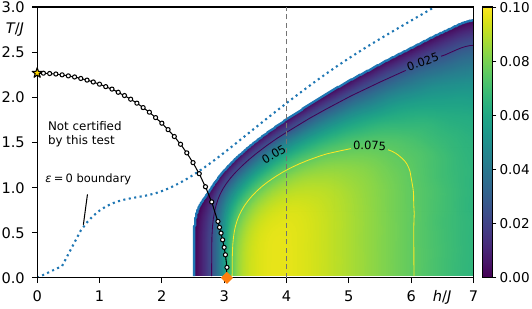}}
  \sbox{\panbD}{\includegraphics{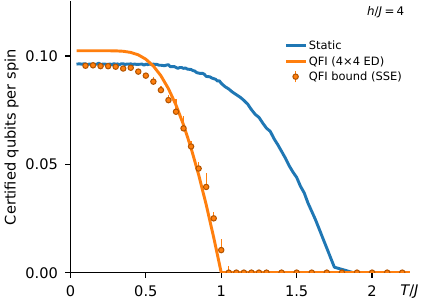}}
  \pgfmathsetmacro{\s}{(\linewidth-\gap)/(\wd\panbA+\wd\panbB)}
  \pgfmathsetlengthmacro{\wA}{\s*\wd\panbA}\pgfmathsetlengthmacro{\wB}{\s*\wd\panbB}
  \pgfmathsetlengthmacro{\xB}{\wA+\gap}
  \pgfmathsetlengthmacro{\HI}{\s*max(\ht\panbA,\ht\panbB)}
  \pgfmathsetlengthmacro{\HII}{\s*max(\ht\panbC,\ht\panbD)}
  \pgfmathsetlengthmacro{\rowsep}{6mm}
  \pgfmathsetlengthmacro{\yI}{\HII+\rowsep}
  \pgfmathsetlengthmacro{\axL}{17.82pt*\s}\pgfmathsetlengthmacro{\axR}{33.80pt*\s}
  \node[anchor=south west] at (0,\yI)    {\includegraphics[width=\wA]{fig6a_chain_thermal.pdf}};
  \node[anchor=south west] at (\xB,\yI)  {\includegraphics[width=\wB]{fig6b_chain_qfi.pdf}};
  \node[anchor=south west] at (0,0)      {\includegraphics[width=\wA]{fig6c_square_thermal.pdf}};
  \node[anchor=south west] at (\xB,0)    {\includegraphics[width=\wB]{fig6d_square_qfi.pdf}};
  \node[anchor=south west] at (0.5mm,\yI+\HI+0.6mm)        {\figletter{a}};
  \node[anchor=south west] at (\axL,\yI+\HI+0.6mm)         {\figtitle{Thermal magic, 1D}};
  \node[anchor=south west] at (\xB+0.5mm,\yI+\HI+0.6mm)    {\figletter{b}};
  \node[anchor=south west] at (\xB+\axR,\yI+\HI+0.6mm)     {\figtitle{The QFI route, 1D}};
  \node[anchor=south west] at (0.5mm,\HII+0.6mm)           {\figletter{c}};
  \node[anchor=south west] at (\axL,\HII+0.6mm)            {\figtitle{Thermal magic, 2D}};
  \node[anchor=south west] at (\xB+0.5mm,\HII+0.6mm)       {\figletter{d}};
  \node[anchor=south west] at (\xB+\axR,\HII+0.6mm)        {\figtitle{The QFI route, 2D}};
\end{tikzpicture}
\caption{{\bf Experimentally accessible volume law of magic.} 
Criteria involving magnetic one- and two-point correlation functions lower bound the density of non-stabilizer resources needed to prepare a state. 
$\varepsilon$ denotes the state preparation tolerance or error. 
In the heat maps the colour and the contours give the certified frame-robust magic
density $n_{\rm fr}$ at $\varepsilon=0.005$; the solid blue line is the edge of the certified region, the dotted blue line the $\varepsilon=0$ boundary.
Results for the 1d TFIM:
{\panelletter{a}} Heat map of the 
volume law in the $T,h$ plane in the thermodynamic limit at $\varepsilon=0.005$. \hcdiamond{}: $h_c/J=1$.
{\panelletter{b}} Quantum Fisher Information criterion 
 Eq.~(\ref{eq:experimental-qfi}) in orange versus temperature for $h=2J$ computed. 
 The green curve shows a restricted energy-window QFI. 
 The blue solid curve is the static criterion (\ref{eq:experimental-volume-law}). 
Results for the 2d TFIM from SSE with $L=48$:
{\panelletter{c}} Certified frame-robust magic density $n_{\rm fr}$ of the 2D TFIM in the $(h,T)$ plane at $\varepsilon=0.005$. \hcdiamond{}: $h_c/J=3.04438(2)$~\cite{Blote2002}. \onsagerstar{}: Ising $T_c/J\simeq 2.269$.
{\panelletter{d}} The QFI route in 2D at $h=4J$: static criterion
(blue) and QFI criterion (orange) versus temperature; markers: 2$\sigma$-certified SSE ($L=48$) lower bounds on $F$ from imaginary-time correlators, converted to $n_{\rm fr}$, orange line the $4\times4$ torus.}
\label{fig:experiments}
\end{figure}

Inelastic neutron scattering also yields a quantum Fisher information (QFI)
certificate. Define $\mathcal F=\max_{s=\pm1}F_Q[\rho,(\sigma_i^x+s\sigma_j^x)/2]$,
where $F_Q[\rho,A]$ is the full-state QFI for generator
$A$~\cite{BraunsteinCaves1994}; the signs select in-phase or out-of-phase
neighbours. A QFI exceeding the separable bound witnesses multipartite
entanglement~\cite{Hyllus2012,Toth2012} and was used in neutron scattering on quantum materials~\cite{Mathew_2020,mazza2026quantum}; here it witnesses magic, because for this
local pair generator the QFI of any locally rotated stabilizer mixture is bounded by
the same pair inequality as the static certificate (see~\ref{sm:readout}), so an
excess certifies non-stabilizerness and not only entanglement. 
Since $\mathcal F\leq2(1+|C_1|)$,
\begin{equation} 
 2m^2+\mathcal F>4
 \quad\Longrightarrow\quad
 \boxed{n_{\rm fr}\geq2m^2+\mathcal F-4>0.}
 \label{eq:experimental-qfi}
\end{equation}
The equilibrium spectral weights are nonnegative: clean, independently normalised partial spectra lower-bound $\mathcal F$~\cite{Hauke2016,Scheie2021,Scheie2025Tutorial}. Missing magnetic weight only weakens certification; calibration and finite-tolerance extensions are in Secs.~\ref{sm:readout} and~\ref{sm:uncertainty}. Crucially, when the above inequalities (\ref{eq:experimental-volume-law}) and (\ref{eq:experimental-qfi}) are not satisfied, we cannot claim the absence of magic: it is a failure of the specific criteria. 

The static certificate survives heating and finite preparation tolerance (Fig.~\ref{fig:experiments}a,b). For the 1d TFIM, the observables are the ones given above Eq.~(\ref{eq:experimental-volume-law})~\cite{Barouch1971}. With state preparation tolerance or error $\varepsilon=0.005$ in trace distance, optimized bounds exceed $0.02$ qubits per spin at $h=0.9J$, $T=0.2J$, and $0.21$ at $h=2J$, $T=0.5J$. The latter still certifies $>0.13$ per spin with simultaneous uncertainties of $\pm0.01$ per observable, in addition to that tolerance. This is an illustrative budget, not achieved precision: we retain only what \emph{every} state consistent with the calibrated uncertainties must contain.
The response route is also nontrivial: on a 16-spin Ising ring at $h=2J$, $T=0.5J$ and the same $\varepsilon$, the QFI bound exceeds $0.13$ per spin using only the energy-window $2J\leq E\leq6J$ (Fig.~\ref{fig:experiments}b). 

Using stochastic series expansions~\cite{Sandvik1992,Sandvik2003} (SSE) it is possible to consider the 2d square lattice TFIM which has a finite temperature $T_c$ phase boundary starting at the Ising limit $T_c/J\simeq 2.269$
at $h=0$, terminating at the critical field $h_c/J=3.04438(2)$~\cite{Blote2002}. 
From SSE on $48\times 48$ lattices, we evaluate the different terms in Eq.~(\ref{eq:experimental-volume-law}), 
obtaining a lower bound on $n_{fr}$. Results for the square lattice TFIM are shown in Fig.~\ref{fig:experiments}c. Although $n_{fr}$ is reduced with respect to the 1d results, it is remarkably similar, with the presence of the finite $T_c$ having negligible influence.
Significant frame-robust magic is certified at temperatures at the same scale as $T_c$. At large $h/J$ the frame-robust magic extends to even higher temperatures.
At $h/J=4$ we use SSE to explore the QFI route,
Fig.~\ref{fig:experiments}d along with $4\times 4$ exact diagonalization results.

\section{Conclusion}
We established sudden death thresholds for quantum magic during typical thermal and dynamical evolutions. Based on the geometry of non-magical states, we have argued that typical trajectories first lose magic, then entanglement, which we confirmed with quenches in spin chains. For thermal evolution, we found that the order is inverted: entanglement typically dies before magic in Gibbs states of local Hamiltonians. In order to support our claims, we developed a powerful algorithm to evaluate a true magic monotone in generic mixed states: Stabilizer Orbit Frank--Wolfe Algorithm (SOFA).
For example, SOFA allowed us to precisely determine the temperature at which all magic dies in the 1d TFIM in the thermodynamic limit. We also introduced a powerful notion of relative magic between the parties. In the groundstate of the TFIM, we showed that it suffers a sudden death with separation strictly before entanglement. Finally, we showed that two bulk measurements
reveal magic that no independent rotations of the spins can remove. When this local signature persists throughout homogeneous matter, it demands a
non-stabilizer preparation resource proportional to the number of spins, even for thermal mixed states, in any spatial dimension.
Given the universality of our conclusions and methods, they can be used to chart the fate of magic in a plethora of quantum many-body systems, ranging from Gibbs states of Hamiltonians to quantum circuits, both in simulations and the laboratory. 

\section{Methods}
{\bf Stabilizer membership.} STAB membership was established constructively, rather than inferred from a small optimized distance. We expressed each reduced density matrix in the Pauli basis and solved sparse linear programmes over adaptively generated pools of stabilizer states, exploiting verified target symmetries and reusing candidate states between neighbouring points. To obtain interior certificates, we sought a stabilizer decomposition of the radially extended state $\rho_\alpha=I/D+\alpha(\rho-I/D)$, where $D$ is the subsystem dimension. A feasible solution with $\alpha>1$ yields a decomposition of $\rho$ with a positive maximally mixed contribution. The resulting weights were independently checked for non-negativity, normalization and reconstruction of all Pauli expectations within the prescribed numerical tolerances. Inconclusive searches were retained as unresolved.
Details of the method can be found in Supplemental Materials~\ref{SM:stab_membership}. 

{\bf Certifying magic.} Let $\sigma=\sum_S p_S\ketbra{S}{S}$ be a stabilizer mixture close to $\rho$, which can be obtained via SOFA or equivalent methods with Hilbert-Schmidt metric. The residual $W=\rho-\sigma$ then serves as a witness: $\rho\notin\mathrm{STAB}$ whenever
\begin{equation}
\operatorname{Tr}(W\rho)>\max_{S}\,\langle S|W|S\rangle ,
\end{equation}
where the maximum runs over all pure stabilizer states $\ket{S}$. The criterion holds for any Hermitian operator $W$, so an inexact projection cannot produce a false certificate. We evaluate the maximum exactly by exhaustive enumeration; since the number of pure stabilizer states grows as $2^{L^2/2+O(L)}$, this is feasible only up to $L=7$.

{\bf Sampling the boundary of STAB.} For $n$-qubit states with dimension $D=2^n$, and $K$-dimensional environment we draw a complex Ginibre rectangular matrix $G\in \mathbb C^{D\times K}$~\cite{Karol_Zyczkowski_2001}.
We then form a density matrix $\rho=G G^\dag/\Tr (GG^\dag)$. 
This is equivalent to drawing a Haar-random pure state in $\mathbb C^D\otimes\mathbb C^K$ and tracing out the $K$-dimensional environment. We work with $n=6$ qubits, and choose 100 independent targets at $K\in \{ 96, 144, 192, 208\}$. When $K>2^6$ the targets are almost surely full rank. The  goal is to project each target $\rho$ onto a boundary state $\sigma_\star\in\partial{\rm STAB}$ via the Bures distance:
\begin{align}
	\min_{\sigma\in \mbox{STAB}} d_B^2(\rho,\sigma)= d_B^2(\rho,\sigma_\star)
\end{align}
Since we are using our SOFA algorithm, we get a good approximation $\widehat\sigma$ to $\sigma_\star$. 
The optimizer used the same \emph{exhaustive} 315,057,600-vertex stabilizer LMO as was used to obtain the $L=6$ TFIM results in Fig.~\ref{fig:ising}. The global Frank--Wolfe gap and strong convexity of the full-rank Bures objective give a certified Hilbert-Schmidt radius $R$ with $||\widehat\sigma-\sigma_\star||_F\leq R$. Define $\Gamma_S$ as the partial transpose with respect to a given bipartite cut $S$. 
For all bipartite cuts $S$, Weyl bounds transfer the iterate eigenvalues to the unknown exact minimizer. If $\min_S \lambda_{\rm min}(\widehat\sigma^{\Gamma_S})<-R$ then $\sigma_\star$ is NPT; if it is $>R$, the projection is certified all-cut PPT. Likewise $\lambda_{\rm min}(\widehat \sigma)>R$ certifies full rank of $\sigma_\star$. 

All 400 exact Bures projections are certified boundary points, full rank, and PT-resolved. All projections are
rigorously NPT at $K=96, 144, 192$. At $K=208$, 98/100 are rigorously NPT and 2/100 are rigorously all-cut PPT. Thus the
cohort contains 398 NPT and 2 all-cut-PPT STAB boundary points. 

{\bf Thermal and Time--dependent density matrices.} 
For the 1d TFIM the Gibbs state of the infinite chain
is Gaussian in the Jordan--Wigner fermions~\cite{Pfeuty1970,Barouch1971}.
The reduced density matrix (RDM) is then formed from the
Majorana correlation matrix~\cite{Peschel2003}.
The time-dependent RDMs of the integrable quench of Sec.~\ref{sec:Fatewtime} are obtained by the same construction.
The Bell-dimer initial state is Gaussian, so the evolution remains within the Gaussian manifold.
For the Kitaev model the Hamiltonian is quadratic in the itinerant
Majoranas within each $\mathbb{Z}_2$ flux sector~\cite{Kitaev2006}.
The Gibbs state is then a Boltzmann-weighted sum over sectors~\cite{Nasu2014,Nasu2015}, which we evaluate exactly on N-site tori.
The RDMs are then mixtures of Gaussian states. For the generic chains of Eq.~(\ref{eq:H_rnd}), which
admit no free-fermion representation, the Gibbs state of an  open chain is
represented as a purified matrix product state~\cite{Verstraete2004,FeiguinWhite2005},
cooled from $\beta=0$ by imaginary-time evolution with a symmetric second-order
three-site time-evolving block decimation (TEBD) sweep~\cite{Vidal2003}. The thermal RDM of the central block is then obtained from the MPS, by contraction.

\section*{Acknowledgements}
We acknowledge useful discussions with Anthony Stibre, Gilles Parez, Jos\'ephine Potdevin, Marcello Dalmonte. 
W.W.-K.\/ and L.L.\/ are supported by a grant from the Fondation Courtois, a Chair of the Institut Courtois, a Discovery Grant from NSERC, and a Canada Research Chair. L.L.\/ acknowledges support from the University of Toronto Arts \& Science Postdoctoral Fellowship Program. E.S.S. acknowledges the support of the Natural Sciences and Engineering Research Council of Canada (NSERC) through Discovery Grant RGPIN-2024-06711. This research was enabled in part by support provided by SHARCNET (sharcnet.ca) and the Digital Research Alliance of Canada (alliancecan.ca). 

OpenAI Codex (OpenAI) and Claude (Anthropic) were used to assist with drafting and refactoring portions of the research software, with figure preparation, and to improve readability of the manuscript. The generated code was reviewed, tested, and validated by the authors. The authors wrote the majority of the manuscript. The authors take full responsibility for the software, analyses, integrity, and conclusions reported in this work.

While finalizing the present manuscript \cite{Sabharwal2026} appeared on the ArXiv, presenting related results concerning the temperature dependence of magic in the Kitaev spin liquid.

\bibliographystyle{apsrev4-2}
\bibliography{ref}

\clearpage
\onecolumngrid

\renewcommand{\thesection}{SM\arabic{section}}
\renewcommand{\theequation}{SM\arabic{equation}}
\renewcommand{\thefigure}{SM\arabic{figure}}
\renewcommand{\thetable}{SM\arabic{table}}
\setcounter{section}{0}
\setcounter{equation}{0}
\setcounter{figure}{0}
\setcounter{table}{0}

\newpage
\begin{center}
{\Large\bfseries Supplementary Information}\par\vspace{3pt}
{\large Supplement to \emph{Fate of quantum magic}}
\end{center}

This supplement establishes the pair-based result used in the main text: a persistent local correlation signature forces a volume law of frame-robust magic. We give the definitions, local bounds and a short counting proof, followed by the experimental hierarchy and the numerical prescriptions needed for the quoted benchmarks. The theorem does not assume independent pairs, an Ising Hamiltonian or a pure state.

\section{Frame-robust magic and its preparation meaning}
\label{sm:resource}

For a pure state on $L$ spins, define
\begin{equation}
 \operatorname{Stab}(\psi)=\{P\in\mathcal P_L:P|\psi\rangle=|\psi\rangle\},
 \qquad \nu(\psi)=L-\log_2|\operatorname{Stab}(\psi)|,
 \label{eq:nullity}
\end{equation}
where $\mathcal P_L$ is the Pauli group, including phases. The stabilizers are commuting Pauli constraints with certain outcome $+1$. If $L-k$ of them are independent, the state lies in a stabilizer code encoding $k$ logical qubits. A Clifford circuit maps it to $|\phi\rangle_k\otimes|0\rangle^{\otimes(L-k)}$: the $k$ logical qubits carry all its non-stabilizer information \cite{Gottesman1997,Beverland2020}. Ordinary nullity is additive, Clifford invariant and non-increasing on pure branches of stabilizer protocols \cite{Beverland2020}.

We remove magic attributable to the choice of individual spin frames before minimizing over ensembles:
\begin{align}
 \nu_{\rm fr}(\psi)&=\min_{U_{\rm loc}=\bigotimes_iU_i}\nu(U_{\rm loc}|\psi\rangle),
 \qquad U_i\in SU(2),\nonumber\\
 \Nfr(\rho)&=\inf_{\rho=\sum_a p_a|\psi_a\rangle\langle\psi_a|}
                   \sum_a p_a\nu_{\rm fr}(\psi_a),
 \qquad \nfr(\rho)=\frac{\Nfr(\rho)}{L}.
 \label{eq:frame-resource}
\end{align}
Each ensemble component chooses its own product of one-spin rotations. This order is essential. Ordinary convex-roof nullity and local-unitary optimization of magic have appeared separately \cite{Tarabunga2025,QianWang2025}; Eq.~\eqref{eq:frame-resource} specifies the component-wise, sitewise construction used here.

Let $\K_L$ be the convex hull of independently locally rotated pure stabilizer states. The quantity $\Nfr$ is convex, locally unitary invariant and bounded between zero and $L$. Its zero set is exactly $\K_L$. Indeed, the total probability of components with nonzero integer $\nu_{\rm fr}$ is at most their average nullity, so a vanishing average places $\rho$ arbitrarily close to the compact set $\K_L$.

\textbf{Preparation-cost lemma.} If the only non-stabilizer input to a protocol is an arbitrary register of at most $k$ qubits, every output satisfies $\Nfr(\rho)\leq k$. Clifford gates, stabilizer ancillas, adaptive Pauli measurements, classical randomness and discarding are allowed. To prove the statement, decompose the input and resolve every measurement outcome. Each pure input branch has ordinary nullity at most $k$, which cannot increase under these operations. Discarding can be represented by averaging Pauli-basis measurement outcomes on the discarded spins. Finally minimize over frames and over the output ensemble \cite{Beverland2020}.

This lemma is the operational claim. Arbitrary local rotations are not additional free gates that may be interleaved with entangling Cliffords. In fact, an entangling Clifford can increase $\nu_{\rm fr}$ by converting removable one-spin magic into correlations. The frame-robust quantity is a preparation-cost lower bound, not an unrestricted Clifford monotone.

\section{Local correlations and the enlarged stabilizer null set}
\label{sm:local}

On a pair $e=(i,j)$ choose orthogonal calibrated Pauli axes at each spin and define
\begin{equation}
 m_e=\frac{\langle\sigma_i^z\rangle+\langle\sigma_j^z\rangle}{2},
 \quad C_{\alpha\alpha,e}=\langle\sigma_i^\alpha\sigma_j^\alpha\rangle,
 \quad C_{{\rm eff},e}=\frac{1-C_{zz,e}+|C_{xx,e}-C_{yy,e}|}{2}.
 \label{eq:coordinates}
\end{equation}
Equivalent sites give $m_e=m$ of the main text. Correlations are full equal-time expectations, not connected correlators. Every pure two-qubit stabilizer is product or maximally entangled. Its local-unitary orbit therefore gives all pure states of the corresponding type, so $\K_2$ is exactly their convex hull.

\textbf{Local all-frame bound.} Every $\tau\in\K_2$ obeys $m_e^2+C_{{\rm eff},e}\leq1$. For a product state put $a=\langle\sigma_i^z\rangle$, $b=\langle\sigma_j^z\rangle$. The Bloch-vector constraints imply
\begin{equation}
 |C_{xx,e}-C_{yy,e}|\leq\sqrt{(1-a^2)(1-b^2)}
 \leq1-\frac{a^2+b^2}{2},
 \qquad C_{{\rm eff},e}\leq1-\frac{(a+b)^2}{4}.
 \label{eq:product-proof}
\end{equation}
For a maximally entangled pure state $m_e=0$. The operators
\begin{equation}
 Q_\eta=\frac{I-\sigma_i^z\sigma_j^z+
                  \eta(\sigma_i^x\sigma_j^x-\sigma_i^y\sigma_j^y)}{2},
 \qquad \eta=\pm1,
 \label{eq:Q}
\end{equation}
each have eigenvalues $\{1,1,1,-1\}$. Hence $C_{{\rm eff},e}=\max_\eta\Tr(Q_\eta\tau)\leq1$. Convexity of $m_e^2+C_{{\rm eff},e}$ extends the bound to all of $\K_2$.

\textbf{Why the extra polarization information helps.} Positivity of the two Bell projectors $[I-\sigma_i^z\sigma_j^z\pm(\sigma_i^x\sigma_j^x+\sigma_i^y\sigma_j^y)]/4$ gives $|C_{xx,e}+C_{yy,e}|\leq1-C_{zz,e}$. Therefore
\begin{equation}
 2|C_{xx,e}|\leq |C_{xx,e}-C_{yy,e}|+|C_{xx,e}+C_{yy,e}|
 \leq2C_{{\rm eff},e}.
 \label{eq:dominance}
\end{equation}
The same statement holds with $x,y$ interchanged. Thus $|C_{xx}|\leq\Ceff$ for every physical pair, not just Ising states. The economical separability criterion $m^2+|C_{xx}|\leq1$ is a two-spin specialization of spin squeezing \cite{sorensen}; the argument above establishes the larger null set and the polarization-resolved improvement.

\textbf{Converting a violation into a local weight.} Define
\begin{equation}
 w_2(\tau)=\min\{w:\tau=(1-w)\sigma+w\omega,\ 
 \sigma\in\K_2,\ \omega\text{ a state},\ 0\leq w\leq1\}.
 \label{eq:weight}
\end{equation}
This convex, locally invariant weight vanishes on $\K_2$; it is not a distance. The eigenvalues of $\lambda(\sigma_i^z+\sigma_j^z)+Q_\eta$ are $-q,1,1,q$, with $q=\sqrt{1+4\lambda^2}$. Maximizing over the sign of $m_e$ and using $m_e^2=\sup_{\lambda\geq0}(2\lambda|m_e|-\lambda^2)$ shows that
\begin{equation}
 m_e^2+C_{{\rm eff},e}-1\leq
 \sup_{\lambda\geq0}\bigl(\sqrt{1+4\lambda^2}-1-\lambda^2\bigr)=\frac14.
 \label{eq:quantum-maximum}
\end{equation}
A free component contributes at most zero, and a residual component at most $1/4$. Convexity therefore gives
\begin{equation}
 w_2(\rho_e)\geq4(m_e^2+\underline C_e-1)
 \quad\text{for every certified }\underline C_e\leq C_{{\rm eff},e}.
 \label{eq:local-weight-bound}
\end{equation}
A negative right-hand side is valid but uninformative. A violation in one calibrated frame suffices; detecting a state in every frame is not required.

\section{From a local signal to a volume law}
\label{sm:volume-law}

The central idea is \emph{no double counting}: distinct nonfree pairs on disjoint spins require distinct logical qubits. Only the pair case needed for the experiment is established here.

\textbf{Pair-packing proposition.} For any $L$-spin state and any collection $\mathcal M$ of mutually disjoint pairs,
\begin{equation}
 \Nfr(\rho)\geq\sum_{e\in\mathcal M}w_2(\rho_e).
 \label{eq:packing}
\end{equation}

\emph{Counting argument for a pure state.} Suppose the ordinary nullity is $k$, so its stabilizer code has $k$ logical qubits. In a pair's Pauli expansion, any term anticommuting with a stabilizer has zero expectation. If all the surviving supported Paulis commute, the reduction is diagonal in a common Pauli eigenbasis and hence is a stabilizer mixture. A nonfree reduction therefore needs two anticommuting supported Paulis that commute with the stabilizer group: an anticommuting logical pair.

Supported Paulis on disjoint physical pairs commute with one another. Their anticommuting logical pairs must consequently be independent. Explicitly, if a product of the chosen logical operators were a stabilizer, commuting it past the partner on one selected pair would force that operator's exponent to vanish; applying the argument to each partner removes every exponent. There are only $2k$ independent logical Pauli directions in a code encoding $k$ qubits \cite{Gottesman1997}. If $q$ disjoint reductions are nonfree, this gives $2q\leq2k$. Since $0\leq w_2\leq1$ and free reductions have zero weight, $\sum_e w_2(\rho_e)\leq q\leq k$.

\emph{Frames and mixtures.} Applying the preceding argument after any product of local rotations leaves each $w_2$ unchanged. Minimization over frames gives $\sum_e w_2\leq\nu_{\rm fr}$ on a pure branch. Convexity of $w_2$ then bounds the sum for a mixed state by the ensemble average of $\nu_{\rm fr}$. Minimizing that average proves Eq.~\eqref{eq:packing}. This proof does not assume that the reductions are independent.\hfill$\square$

\textbf{The bulk consequence.} In a Bravais crystal, links at one fixed nonzero neighbour displacement form parallel paths. Taking alternate links along each path pairs all but a subextensive boundary set: $M_L=|\mathcal M|=L/2-o(L)$ for regular growing regions. This pairing is a proof device, not a dimerization of the Hamiltonian. If the local measured margin repeats in the bulk, Eqs.~\eqref{eq:local-weight-bound} and \eqref{eq:packing} give
\begin{equation}
 \Nfr(\rho)\geq4M_L(m^2+\underline C-1),\qquad
 \nfr:=\liminf_{L\to\infty}\frac{\Nfr(\rho)}L
 \geq2(m^2+\underline C-1).
 \label{eq:volume-law}
\end{equation}
A size-independent positive margin supplies a linear lower bound; $\Nfr\leq L$ supplies the upper bound. Therefore $\Nfr=\Theta(L)$: \emph{a volume law is proved, not inferred by fitting finite-size data}. The number $L$ counts spins, not linear size, and no purity or spatial-dimension assumption enters the counting argument.

For a different resolved neighbour family, let $\gamma=\liminf M_L/L>0$ be its disjoint-pair density; the coefficient $2$ is replaced by $4\gamma$. The Bravais construction is a sufficient physical setting for $\gamma=1/2$, not a requirement that every spin be individually measured. A single isolated violation, without an extensive number of equivalent links, does not by itself establish a volume law.

\section{Static and dynamical experimental hierarchy}
\label{sm:readout}

\textbf{Static readout.} Within a calibrated effective doublet, write $\bm\mu_i=\mu_B\bm g_i\bm\sigma_i/2+\bm\mu_i^{\rm VV}$. For equivalent sites on a principal axis, magnetometry gives $m=2M_z/(L\mu_Bg_z)$ after subtracting non-spin backgrounds. Let $S^{\alpha\alpha}(\bm q,E)$ be the magnetic structure factor per site for $S^\alpha=\sigma^\alpha/2$. Its normalization and the neighbour Fourier moment are
\begin{align}
 \frac1{V_{\rm BZ}}\int_{\rm BZ}\dd^dq\int_{-\infty}^{\infty}\dd E\,
 S^{\alpha\alpha}(\bm q,E)&=\frac14,\nonumber\\
 C_{\alpha\alpha}&=\frac4{V_{\rm BZ}}\int_{\rm BZ}\dd^dq\,
 \cos(\bm q\cdot\bm\delta)\int_{-\infty}^{\infty}\dd E\,
 S^{\alpha\alpha}(\bm q,E).
 \label{eq:static-readout}
\end{align}
The energy integral includes elastic, quasielastic and inelastic magnetic weight. Absolute normalization, or a validated full-channel sum rule, fixes the intensity scale \cite{Xu2013,Scheie2021,Laurell2025,Scheie2025Tutorial}. A cosine-moment/total-weight ratio cancels one common unknown gain only when the full channel weight is measured or bounded. Form factors, polarization factors, backgrounds and missing coverage remain part of the reconstruction \cite{Moon1969,Scheie2025Tutorial}.

On qubit arrays, $z$-basis samples provide $m$ and $C_{zz}$, $x$-basis samples provide $C_{xx}$, and an additional $y$ setting supplies $C_{yy}$. All settings must sample the same state. A common local rotation of the axes does not invalidate the test; loss of orthogonality, leakage and incorrect outcome assignment still require calibration.

\textbf{Full-state QFI.} For $\rho=\sum_a p_a|a\rangle\langle a|$, use
\begin{equation}
 F_Q[\rho,A]=2\sum_{a,b:p_a+p_b>0}
 \frac{(p_a-p_b)^2}{p_a+p_b}|\langle a|A|b\rangle|^2,
 \qquad F_Q[\rho,A]\leq4\Var_\rho(A).
 \label{eq:qfi-definition}
\end{equation}
Put $A_{s,\alpha}=(\sigma_i^\alpha+s\sigma_j^\alpha)/2$, $F_{s,\alpha}=F_Q[\rho,A_{s,\alpha}]$, and $\mathcal F=\max_{s=\pm1}F_{s,x}$. These are QFIs of the full responding state, not of its pair reduction. Since $A_{s,\alpha}^2=(I+s\sigma_i^\alpha\sigma_j^\alpha)/2$,
\begin{equation}
 F_{s,\alpha}\leq2(1+sC_{\alpha\alpha})
 -(\langle\sigma_i^\alpha\rangle+s\langle\sigma_j^\alpha\rangle)^2
 \leq2(1+sC_{\alpha\alpha}).
 \label{eq:qfi-static}
\end{equation}
For a thermal equilibrium response, with $\hbar=1$ and susceptibility per site for $S^\alpha=\sigma^\alpha/2$ \cite{Hauke2016,Scheie2025Tutorial},
\begin{equation}
 F_{s,\alpha}=\frac4{\pi V_{\rm BZ}}\int_0^\infty\dd E\,
 \tanh\!\left(\frac{E}{2k_BT}\right)
 \int_{\rm BZ}\dd^dq\,2[1+s\cos(\bm q\cdot\bm\delta)]\,
 \chi_{\alpha\alpha}''(\bm q,E).
 \label{eq:qfi-spectrum}
\end{equation}
All factors are nonnegative at $E>0$. Restricting the integral to a clean measured domain $\Omega$ gives $F_{s,\alpha,\Omega}\leq F_{s,\alpha}$. With $\mathcal F_\Omega=\max_s F_{s,x,\Omega}$, the main-text hierarchy follows:
\begin{equation}
 \frac{\mathcal F_\Omega}{2}-1\leq\frac{\mathcal F}{2}-1
 \leq |C_{xx}|\leq\Ceff.
 \label{eq:hierarchy}
\end{equation}
Every rung supplies a valid $\underline C$ in Eq.~\eqref{eq:volume-law}; in particular $\nfr\geq2m^2+\mathcal F-4$ when positive. The QFI advantage is conservative partial coverage, not greater intrinsic thermal sensitivity. The variance inequality is generally strict at finite temperature, and a vanishing resource bound need not mean vanishing QFI. At positive temperature strictly elastic weight has zero QFI weight.

The retained spectrum needs an independently justified scale: renormalizing a partial window to exhaust the spin sum rule destroys conservativeness. The static cosine kernel changes sign, so its missing contributions require separate bounds. Multiple magnetic sites per cell require the sublattice response matrix. If the responding equilibrium system includes an environment, use $F_Q[\varrho,A\otimes I]$; its variance is the spin-state variance, so the same bound follows for calibrated spin factors without identifying this with reduced-state QFI.

\textbf{Additional polarization channels.} The bounds $l_\alpha=F_{+,\alpha}/2-1\leq C_{\alpha\alpha}\leq u_\alpha=1-F_{-,\alpha}/2$ give
\begin{equation}
 \Ceff\geq\frac{1-u_z+\max\{0,l_x-u_y,l_y-u_x\}}2.
 \label{eq:polarization-qfi}
\end{equation}
Partial-spectrum lower bounds can replace each QFI. Retain the largest of this and the earlier valid bounds. Extra information strengthens the inference when existing constraints are retained; distinct polarization formulas need not be individually ordered.

\section{Preparation tolerance and experimental uncertainty}
\label{sm:uncertainty}

Allow a preparation to choose any nearby state of the entire $L$-spin region, and count the least resource it could require:
\begin{equation}
 \Nfr^\varepsilon(\rho)=\inf_{D(\rho,\omega)\leq\varepsilon}\Nfr(\omega),
 \qquad D(\rho,\omega)=\frac12\|\rho-\omega\|_1.
 \label{eq:smoothing}
\end{equation}
The tolerance is held fixed as $L$ grows, not allocated independently per bond. This standard distance-ball construction \cite{Mora2008} preserves the operational exclusion: producing any admissible $\omega$ with only $k$ non-stabilizer input qubits implies $\Nfr^\varepsilon(\rho)\leq k$.

Contractivity of trace distance gives $D(\rho_e,\omega_e)\leq\varepsilon$ on every pair. Apply Eq.~\eqref{eq:packing} to each admissible $\omega$ and take the infimum. With $w_2^\varepsilon(\tau)=\inf_{D(\tau,\zeta)\leq\varepsilon}w_2(\zeta)$ this yields
\begin{equation}
 \Nfr^\varepsilon(\rho)\geq\sum_{e\in\mathcal M}w_2^\varepsilon(\rho_e).
 \label{eq:smoothed-packing}
\end{equation}
No homogeneity of $\omega$ or independence of deviations is assumed.

\textbf{Stability lemma.} A strict measured margin protects all nearby states. The pair magnetization operator has norm one; $\Ceff$ is a maximum of expectations of the norm-one $Q_\eta$. Both quantities change by at most $2\varepsilon$, and the squared magnetization changes by at most $4\varepsilon$. Equations~\eqref{eq:local-weight-bound} and \eqref{eq:smoothed-packing} therefore imply
\begin{equation}
 \nfr^\varepsilon\geq2(m^2+\underline C-1-6\varepsilon)
 \quad\text{when the right-hand side is positive}.
 \label{eq:stable-volume-law}
\end{equation}
Here $\nfr^\varepsilon$ denotes the bulk lower limit per spin and $\underline C\leq\Ceff$ can be any rung of Eq.~\eqref{eq:hierarchy}. For QFI the penalty is applied only after conversion to a bounded static observable; no continuity bound on QFI is assumed. The condition $\varepsilon<(m^2+\underline C-1)/6$ guarantees a positive density, not the unchanged exact-preparation density. Tighter bounds used for the numerical examples are specified in \smref{sm:benchmarks}.

\textbf{The resource required by uncertain data.} Minimize each certificate over the joint confidence region for the observables and their shared calibration parameters. For physically normalised estimates $\widehat m,\widehat c$ of $m,C_{xx}$ with simultaneous absolute uncertainties $u_m,u_c$, a simple sufficient result is
\begin{equation}
 \nfr^\varepsilon\geq2\bigl(\widehat m^{\,2}+|\widehat c|-1
                         -2u_m-u_c-6\varepsilon\bigr)
 \quad\text{when positive}.
 \label{eq:experimental-uncertainty}
\end{equation}
It follows from $|m^2-\widehat m^{\,2}|\leq2u_m$ and $||c|-|\widehat c||\leq u_c$. For $\Ceff$, a conservative absolute uncertainty is $(u_{zz}+u_{xx}+u_{yy})/2$. Axes, signs or energy windows selected from the data must be included in simultaneous inference.

A separately validated reconstruction with $D(\rho_{\rm true},\widehat\rho)\leq\varepsilon$ gives $\Nfr(\rho_{\rm true})\geq\Nfr^\varepsilon(\widehat\rho)$. However, uncertainty on two expectations does not establish a whole-state trace-distance ball. Preparation tolerance and measurement uncertainty can enter the same bound without being identified with each other.

\section{TFIM benchmarks and numerical prescription}
\label{sm:benchmarks}

\textbf{Exactly solvable static testbed.} The convention throughout is
\begin{equation}
 H=-J\sum_j\sigma_j^x\sigma_{j+1}^x-h\sum_j\sigma_j^z,
 \qquad g=h/J.
 \label{eq:tfim}
\end{equation}
For the symmetry-preserving thermodynamic equilibrium state \cite{Pfeuty1970,Barouch1971},
\begin{align}
 \Lambda_k&=\sqrt{1+g^2-2g\cos k},\nonumber\\
 G_r(T)&=\int_0^\pi\frac{\dd k}{\pi}
 \frac{\cos(rk)(\cos k-g)-\sin(rk)\sin k}{\Lambda_k}
 \tanh\!\left(\frac{J\Lambda_k}{k_BT}\right),\nonumber\\
 m_z&=-G_0,\quad C_{xx}=G_{-1},\quad C_{yy}=G_1,
 \quad C_{zz}=m_z^2-G_{-1}G_1.
 \label{eq:tfim-correlators}
\end{align}
At $T=0$ the thermal factor is one. At $g=1$, $m_z=C_{xx}=2/\pi$ and $C_{yy}=-2/(3\pi)$. The minimal-channel certificate at that field is positive up to $k_BT/J\simeq0.513253$. This is a certification threshold, not a temperature at which all magic disappears.

\textbf{Local inference used for the quoted values.} The finite-tolerance benchmarks enforce physical pair density matrices rather than only the universal penalty in Eq.~\eqref{eq:stable-volume-law}. For $M=(\sigma_i^z+\sigma_j^z)/2$ and $X=\sigma_i^x\sigma_j^x$, define
\begin{equation}
 G_\varepsilon(m,c)=\inf\{w_2(\omega):\tau,\omega\text{ states},\ 
 \Tr(M\tau)=m,\ \Tr(X\tau)=c,\ D(\tau,\omega)\leq\varepsilon\}.
 \label{eq:local-inference}
\end{equation}
The actual reduction is one admissible $\tau$; hence Eq.~\eqref{eq:smoothed-packing} gives $\nfr^\varepsilon\geq G_\varepsilon(m,c)/2$. Only measured moments constrain $\tau$. Resource inference from incomplete data is an established strategy \cite{Guhne2007}.

The small optimization is fully specified by Eq.~\eqref{eq:local-inference} and the free-cone identity
\begin{equation}
 w_2(\omega)=\min\{\Tr R:\omega=S+B+R,\ S,B,R\geq0,\ 
 S^{T_j}\geq0,\ \Tr_iB=\Tr_jB=(\Tr B)I_2/2\}.
 \label{eq:free-cone}
\end{equation}
In two qubits, positive partial transpose is equivalent to separability \cite{Horodecki1996PPT}. A state with both marginals maximally mixed is locally Bell diagonal and therefore a mixture of maximally entangled states \cite{Horodecki1996States}. These two facts prove Eq.~\eqref{eq:free-cone}. This compact prescription is included because the main-text numbers depend on it; a wider optimization theory is not needed here.

\begin{table}[ht]
\centering
\caption{Minimal-channel density lower bounds at preparation tolerance $\varepsilon=0.005$, retained from the main-figure calculations. Static rows use thermodynamic correlations. The last two rows use the full Gibbs state of a 16-spin periodic ring and the same magnetization. Entries are rounded downward, not exact resource values.}
\begin{tabular}{@{}lccc@{}}
\toprule
Information used & $h/J$ & $k_BT/J$ & Certified qubits per spin\\
\midrule
Static, $m_z,C_{xx}$ & 0.8 & 0.10 & 0.00768\\
Static, $m_z,C_{xx}$ & 0.9 & 0.20 & 0.02072\\
Static, $m_z,C_{xx}$ & 1.0 & 0.20 & 0.05785\\
Static, $m_z,C_{xx}$ & 2.0 & 0.50 & 0.21193\\
Full QFI and $m_z$, 16 spins & 2.0 & 0.50 & 0.13916\\
E-window QFI and $m_z$, 16 spins & 2.0 & 0.50 & 0.13130\\
\bottomrule
\end{tabular}
\label{tab:benchmarks}
\end{table}

At $h=2J$, $k_BT=0.5J$, thermodynamic quadrature gives $m_z=0.9287782102$, $C_{xx}=0.2629565782$ and margin $0.1255855419$. The illustrative simultaneous uncertainties $u_m=u_c=0.01$, together with $\varepsilon=0.005$, give $\nfr^\varepsilon>0.13117$ directly from Eq.~\eqref{eq:experimental-uncertainty}. This is a sufficient uncertainty budget, not an achieved instrumental precision.

\textbf{Finite-ring dynamical benchmark.} Panel b of the main figure uses the full 16-spin Gibbs state at $h=2J$. All $2^{16}$ eigenstates, both parity sectors and all lattice momenta enter Eq.~\eqref{eq:qfi-definition}. The E-window retains $2J\leq|E_b-E_a|\leq6J$ without renormalization; the full momentum sum is retained. At $k_BT=0.5J$, the stored values are $F_{+,x}=2.45438680$ and $F_{+,x,\Omega}=2.44663294$.

For the QFI inference, replace the equality on $c$ in Eq.~\eqref{eq:local-inference} by $c\geq F_{+,x}/2-1$. Equivalently, evaluate $G_\varepsilon$ at this lower value when positive: $G_\varepsilon(m,c)$ is convex and even in $c$, hence nondecreasing for $c\geq0$. Evenness follows by conjugating one spin by $\sigma^z$. The $+$ channel is the larger one in this benchmark; both channels are computed.

The maximum sampled 14-to-16-site change of the QFI-derived density bound is below $1.6\times10^{-4}$ per spin. This checks finite-size sensitivity, not an infinite-chain limit. Only dual or independently justified analytic lower bounds are reported; a primal minimum alone is not a certified lower bound. The saved 16-spin inference has a largest local primal--dual gap below $3.9\times10^{-6}$ per spin. Calculations use floating point, not interval arithmetic. The accompanying numerical records and scripts specify the values and checks. The benchmarks are retained from the main-figure calculation.

All numerical densities in Table~\ref{tab:benchmarks} use the minimal channel or its QFI lower bound, not the enhanced $\Ceff$ rung. The theorem is algebraic and independent of these model calculations. Neither a zero plotted bound nor loss of a spectral-window signal establishes absence of magic. These are preparation-cost bounds, not distillation rates or guarantees of quantum-computational advantage.

\begin{figure}
    \centering
    \includegraphics[width=0.85\linewidth]{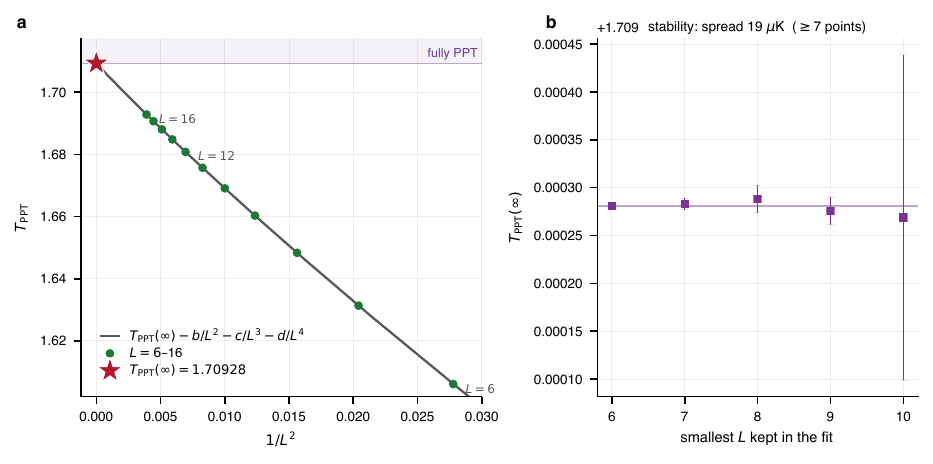}
    \caption{a) PPT temperature for an interval of $L$ sites vs $1/L^2$. b) Stability analysis for the $L\to\infty$ extrapolation. }
    \label{fig:tfim_ppt}
\end{figure}

\section{Certifying Stabilizer Membership}\label{SM:stab_membership}

\begin{figure}[ht!]
    \centering
    \includegraphics[width=0.75\linewidth]{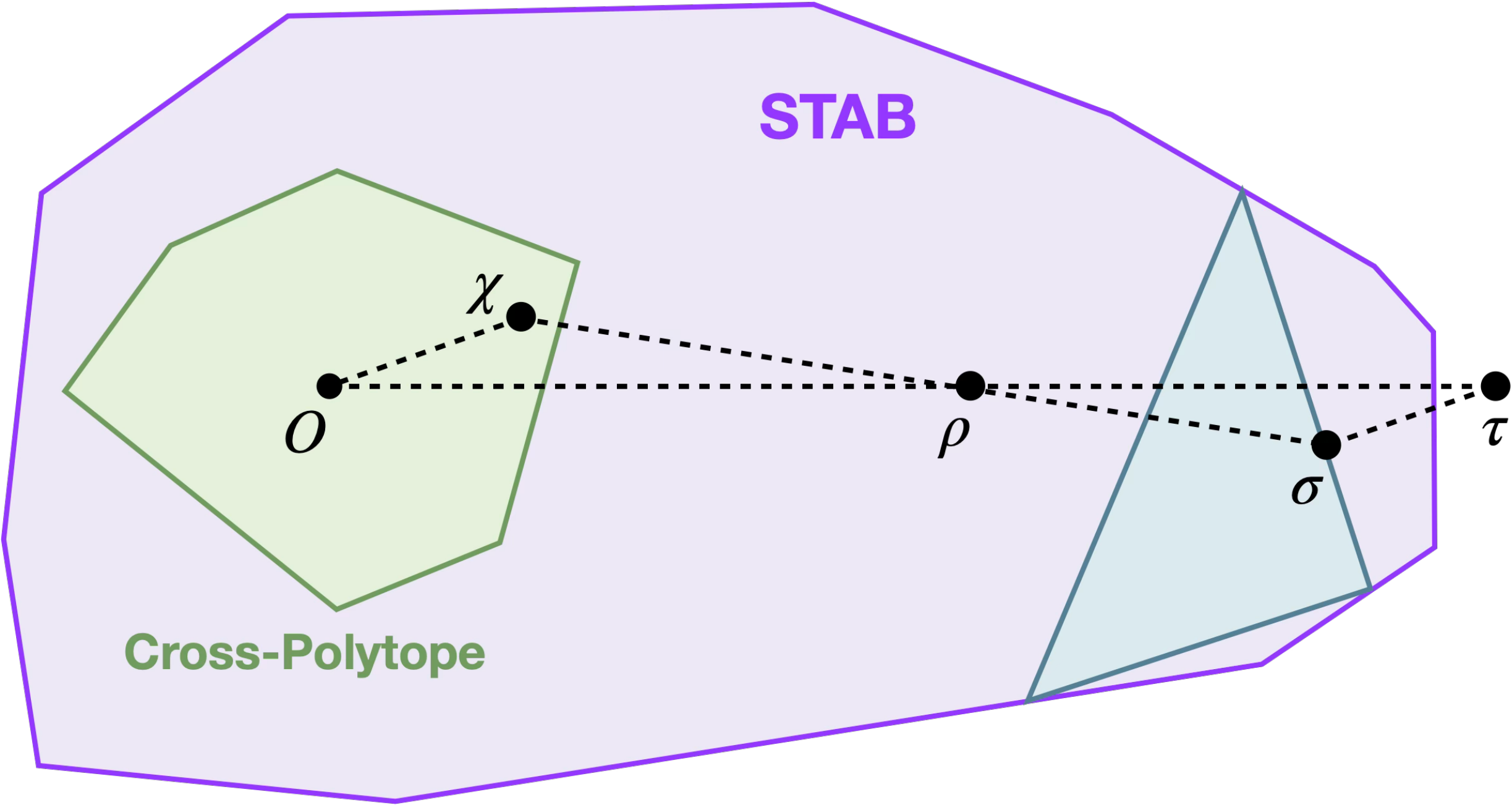}
    \caption{\textbf{Geometry of stabilizer membership certification.}
    The target $\rho$ is inflated radially from $O$ to $\tau$, which is
    approximated by the stabilizer mixture $\sigma$. The residual
    determines $\chi$; the Pauli-$\ell_1$ test places $\chi$ in the
    green cross-polytope inside $\mathrm{STAB}_n$. Equation
    \eqref{eq:stab_convex_identity} then expresses $\rho$ as a convex
    mixture of $\sigma$ and $\chi$. The polygons are schematic.}
    \label{fig:STAB_certificate}
\end{figure}

We introduce a method that certifies that a given $n$-qubit state belongs to the stabilizer polytope (STAB) by constructing an explicit convex decomposition. The method is inspired by a previous work where the Gilbert algorithm~\cite{Gilbert1966,brierley2016Gilbert} was used to certify the absence of entanglement~\cite{Shang2018}. 
Let $\rho$ be an $n$-qubit density matrix, $D=2^n$, and $O=I/D$. The construction is illustrated schematically in Fig.~\ref{fig:STAB_certificate} and proceeds in four steps:
\begin{enumerate}
    \item \emph{Radial inflation.} Choose $\eta>0$ and set
    \begin{equation}
        \tau=O+(1+\eta)(\rho-O)=(1+\eta)\rho-\eta O.
    \end{equation}
    We choose $\eta$ so that $\tau\succeq0$; $\tau$ need not belong to
    $\mathrm{STAB}_n$.

    \item \emph{Stabilizer approximation.} Find a mixture of pure stabilizer
    projectors close to $\tau$,
    \begin{equation}
        \sigma=\sum_{i=1}^{m}w_i\ketbra{\psi_i}{\psi_i},
        \qquad w_i\geq0,\quad\sum_i w_i=1.
    \end{equation}
    The pool and weights may be found, for example, using the stabilizer
    orbit Frank--Wolfe algorithm (SOFA).

    \item \emph{Residual correction.} Define $R=\tau-\sigma$ and the
    candidate correction operator $\chi=O+R/\eta$. Rearranging gives
    \begin{equation}\label{eq:stab_convex_identity}
        \rho=\frac{\sigma+\eta\chi}{1+\eta}.
    \end{equation}
    It therefore remains to certify that $\chi\in\mathrm{STAB}_n$.

    \item \emph{Pauli certificate.} Expand the traceless residual in the
    unsigned Hermitian Pauli basis:
    \begin{equation}
        R=\frac{1}{D}\sum_{P\neq I}\delta_P P,
        \qquad \delta_P=\operatorname{Tr}(PR),
        \qquad \epsilon=\sum_{P\neq I}|\delta_P|.
    \end{equation}
    If $\epsilon\leq\eta$, the decomposition below proves that
    $\chi\in\mathrm{STAB}_n$, and hence, by
    Eq.~\eqref{eq:stab_convex_identity}, $\rho\in\mathrm{STAB}_n$.
\end{enumerate}

To justify the final step, define
$\omega_{P,s}=(I+sP)/D$ for $P\neq I$ and $s=\pm1$.
This is the normalised projector onto the $s$ eigenspace of $P$.
Completing $P$ to a maximal commuting Pauli set shows that
$\omega_{P,s}$ is a uniform mixture of pure stabilizer states. Their
convex hull is the Pauli-$\ell_1$ cross-polytope
\begin{equation}
    \mathcal C_{\mathrm P}
    =\operatorname{conv}\{\omega_{P,s}:P\neq I,\ s=\pm1\}
    =\left\{O+\frac{1}{D}\sum_{P\neq I}x_P P:
                \sum_{P\neq I}|x_P|\leq1\right\}
    \subseteq\mathrm{STAB}_n.
\end{equation}
Indeed, when $\epsilon\leq\eta$,
\begin{equation}\label{eq:chi_stab_decomposition}
    \chi=
    \left(1-\frac{\epsilon}{\eta}\right)O
    +\sum_{\substack{P\neq I\\\delta_P\neq0}}
       \frac{|\delta_P|}{\eta}\,
       \omega_{P,\operatorname{sgn}(\delta_P)}.
\end{equation}
The coefficients are nonnegative and sum to one, while
$O=(\omega_{P,+}+\omega_{P,-})/2$ for any fixed $P\neq I$.
Thus Eq.~\eqref{eq:chi_stab_decomposition} explicitly places $\chi$
inside $\mathcal C_{\mathrm P}$. 

As an alternative to the Pauli-$\ell_1$ cross-polytope test, one may certify
$\chi$ using the centered stabilizer ball of
Ref.~\cite{liu2026geometrytypicalityquantummagic}. Writing
$a_\ast=1/2.181847$, this criterion is
\begin{equation}
    \operatorname{Tr}(\chi^2)
    =\frac{1}{D}+\frac{\|R\|_{\mathrm{HS}}^2}{\eta^2}
    \leq\frac{1}{D-a_\ast}.
\end{equation}
Equivalently, it suffices that
$\|R\|_{\mathrm{HS}}\leq
\eta\sqrt{a_\ast/[D(D-a_\ast)]}$.
Either test establishes $\chi\in\mathrm{STAB}_n$, so the convex-mixture
identity \eqref{eq:stab_convex_identity} certifies $\rho$.

\section{A Lower Bound on QFI from SSE}~\label{SM:SSE}
The Quantum Fisher Information (QFI) criterion is evaluated from SSE as a rigorous, lower bound on
$\mathcal F$. Since $F_Q[\rho,A]\leq4\langle A^2\rangle$,
one has $\mathcal F\leq2(1+|C_1|)$, so Eq.~\eqref{eq:experimental-qfi} follows from
Eq.~\eqref{eq:experimental-volume-law} and any lower bound on $\mathcal F$ certifies. For the Gibbs
state,
\begin{align*}
F_Q&=4\int d\omega\,S(\omega)\tanh\tfrac{\beta\omega}{2}\,(1-e^{-\beta\omega}),\\
G(\tau)&=S_0+\int d\omega\,S(\omega)\bigl[e^{-\tau\omega}+e^{-(\beta-\tau)\omega}\bigr],
\end{align*}
with $S\geq0$ the spectral weight of the generator~\cite{Hauke2016}, $S_0$ its static weight and
$G$ the imaginary-time correlator. Any coefficients $c_k$ with
\begin{equation*}
\begin{aligned}
&\sum_kc_k\bigl[e^{-\tau_k\omega}+e^{-(\beta-\tau_k)\omega}\bigr]\\
&\qquad\leq4\tanh\tfrac{\beta\omega}{2}\,(1-e^{-\beta\omega})\quad\text{for all }\omega
\end{aligned}
\end{equation*}
therefore give $F_Q\geq\sum_kc_kG(\tau_k)$ (the case $\omega=0$ forces $\sum_kc_k\leq0$, so $S_0$ drops out) without analytic continuation, the best
$c_k$ solving a linear program. We measure $G(\tau)$ 
at $23$--$31$ values of $\tau\in[0,\beta/2]$ in
stochastic series expansion (SSE) simulations~\cite{Sandvik1992,Sandvik2003},
where the generator is diagonal. The coefficients are fitted on one half of the bins and evaluated
on the other, and the best of ten candidates (five coefficient boxes, $s=\pm1$) is certified at
$2\sigma$. Against exact diagonalization of
$4\times3$ and $4\times4$ tori the certified bound lies $0.3$--$2.4\%$ below the exact $\mathcal F$
and never above it.  Fig.~\ref{fig:experiments}d shows $L=48$ tori at $h=4J$ ($10^5$ sweeps in $100$ bins per temperature) and is drawn at $\varepsilon=0.005$.

The preparation tolerance $\varepsilon$ is a worst-case penalty (see \ref{sm:uncertainty}):
$\varepsilon$ is the trace distance between the prepared region and the target, so contractivity bounds the
distance of every measured pair by the same $\varepsilon$. The witness
$B_\lambda=\lambda(\sigma^z_i+\sigma^z_j)+\sigma^x_i\sigma^x_j$ has operator norm $q(\lambda)=\sqrt{1+4\lambda^2}$, hence
its expectation in any state within $\varepsilon$ is lowered by at most $2q(\lambda)\varepsilon$, and
\begin{equation*}
n_{\rm fr}^{\varepsilon}\geq\tfrac12\sup_{\lambda>0}f(\lambda)\bigl[2\lambda|m|+|C_1|-b(\lambda)-2q(\lambda)\varepsilon\bigr],
\end{equation*}
with $b=1+\lambda^2$ for $\lambda\leq1$, $b=2\lambda$ for $\lambda\geq1$ and $f=1/(q-b)$. For $\varepsilon=0$ this is
the sharpened form of Eq.~\eqref{eq:experimental-volume-law}. The QFI route uses $|C_1|\geq\mathcal F/2-1$ for the target
\emph{before} the penalty, so no continuity of $\mathcal F$ in trace distance is needed. Statistical errors are
not folded into $\varepsilon$: the plotted SSE values are $2\sigma$ lower bounds.

{
\section{The $\Wit_K$ witness.}\label{SM:Wk}
$\Wit_K$ is defined as follows:
\begin{equation}
 \Wit_{K}:=\sum_{a=x,y,z}|c_a|\le1\ \text{on }\STAB,\qquad
 c_a:=\ev{\sigma^a_i\sigma^a_{j_a}}_{a\text{-bond}}.
\label{eq:wk}
\end{equation}
The energy per site is
$e=-\tfrac12\sum_aJ_ac_a$, so for isotropic couplings $\Wit_K=3c=\frac{2|e|}{J}$. 
Any four-qubit state is fixed by its Pauli expansion,
\begin{equation}
 \rho_\star=\frac1{16}\sum_{P\in\{\mathbb{I},\sigma^x,\sigma^y,\sigma^z\}^{\otimes4}}\ev{P}\,P,
\label{eq:starexp}
\end{equation}
a sum over $4^4=256$ strings, at zero field only four of them survive,
since spin correlations are strictly bond-local in this model~\cite{Baskaran2007}:
$ \ev{P}=0$ unless $P\in\{\mathbb{I},K_x,K_y,K_z\}.$
It follows that the star reduced density matrix is
exactly $\rho_\star=\tfrac1{16}(\mathbb{I}+\sum_ac_aK_a)$. If $\sum_a|c_a|\le1$
then $\rho_\star=\sum_a|c_a|\,\tfrac{\mathbb{I}+\operatorname{sgn}(c_a)K_a}{16}
+(1-\sum_a|c_a|)\tfrac{\id}{16}$ is a mixture of normalised stabilizer-code
projectors, $\Pi_a^{\pm}:=\frac{\mathbb{I}\pm K_a}{2}$, each of which is a uniform mixture of stabilizer states; if
$\sum_a|c_a|>1$ the witness excludes $\STAB$. Hence,
\begin{equation}
 \rho_\star\in\STAB\iff\sum_a|c_a|\le1\ .
\end{equation}

\section{Two Frame Dependent Witnesses}~\label{SM:frame-witness}
\begin{figure}[h!]
    \centering
    \includegraphics[width=0.75\linewidth]{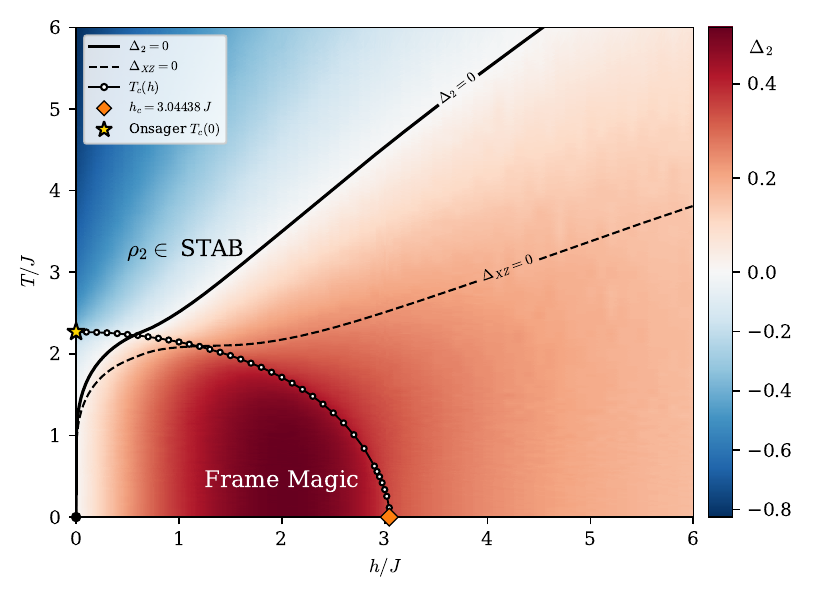}
    \caption{The two-site tests $\Delta_2$ and $\Delta_{XZ}$ in the same plane of the 2D TFIM. The coloring represents $\Delta_2$.  \hcdiamond{}: $h_c/J=3.04438(2)$~\cite{Blote2002}. \onsagerstar{}: Ising $T_c/J\simeq 2.269$. Results are from SSE calculations on a square $48\times 48$ lattice.}
    \label{fig:DeltaTFIM}
\end{figure}
If the local Pauli frame is known, frame dependent witnesses can be considered. The simplest examples
involve only two sites, and on-site and two-site thermal expectation values which are readily
available in experiments. Here we consider two such witnesses, $\Delta_{XZ}=\langle\sigma^z_i\rangle+\langle\sigma^x_i\sigma^x_j\rangle-1$ and the more complete $\Delta_2=\big|\langle\sigma^z_i\rangle+\langle\sigma^z_j\rangle\big|-\langle\sigma^z_i\sigma^z_j\rangle
+\big|\langle\sigma^y_i\sigma^y_j\rangle-\langle\sigma^x_i\sigma^x_j\rangle\big|-1$.
Both are model agnostic in the sense that they witness magic for any thermal many-body $\rho_2$ independently of
any known Hamiltonian or model. 

A pure stabilizer state is the common $+1$ eigenstate of an Abelian group $S$ of Pauli strings with
$-\mathbb{I}\notin S$~\cite{Gottesman1998,Aaronson2004}. Hence, for every Pauli string $P$ we have
$\langle P\rangle=\pm1$ if $\pm P\in S$ and $\langle P\rangle=0$ otherwise. Two anticommuting strings
cannot both lie in an Abelian group, so they cannot both have a non-zero expectation value.

The $\Delta_{XZ}$ witness:
The strings $\sigma^z_i$ and $\sigma^x_i\sigma^x_j$ anticommute. On a pure stabilizer
state at most one of $\langle\sigma^z_i\rangle$ and $\langle\sigma^x_i\sigma^x_j\rangle$ is non-zero, and it
equals $\pm1$, so $|\langle\sigma^z_i\rangle|+|\langle\sigma^x_i\sigma^x_j\rangle|\le1$. The left-hand side is a
convex function of the state, so the bound holds on every mixture of stabilizer states, that is, on all of STAB.
Therefore
$\Delta_{XZ}=\langle\sigma^z_i\rangle+\langle\sigma^x_i\sigma^x_j\rangle-1\le0$ on STAB,
and $\Delta_{XZ}>0$ certifies magic of the full state.

The $\Delta_2$ witness:
Using $|a|=\max_{s=\pm1}sa$, $\Delta_2+1$ is the largest of the four linear functions
$s_1(\langle\sigma^z_i\rangle+\langle\sigma^z_j\rangle)-\langle\sigma^z_i\sigma^z_j\rangle
+s_2(\langle\sigma^y_i\sigma^y_j\rangle-\langle\sigma^x_i\sigma^x_j\rangle)$, $s_1,s_2=\pm1$.
Each is largest at a pure stabilizer state, where it depends only on the reduced state of the pair $(i,j)$,
a mixture of the 60 two-qubit stabilizer states (36 product states and 24 entangled ones). For a product state each
Bloch vector points along an axis and $\Delta_2\in\{0,-1\}$. For an entangled one the local expectation values
vanish and the correlation matrix is a signed permutation, which gives $\Delta_2\in\{0,-1,-2\}$. Hence,
$\Delta_2\le0$ on STAB.

For both witnesses $\Delta>0$ certifies magic; $\Delta_{XZ}\le0$ does not exclude it, only $\Delta_2\le0$ certifies STAB
for the corresponding $\rho_2.$ Results for both witnesses for the 2d square lattice TFIM are shown in Fig~\ref{fig:DeltaTFIM} as obtained from SSE calculations on a $48\times 48$ lattice. 
In this case, the frame dependent magic is more concentrated in the ordered phase below $Tc$, in contrast to the frame-robust magic shown
in Fig~\ref{fig:experiments}c of the main paper, that are concentrated above $h_c/J$. Since only $\Delta_2\le0$ certifies STAB, the
$\Delta_2$ witness always certifies magic at higher temperatures than $\Delta_{XZ}$. As $h/J$ is increased it can be shown that the temperature, $T_W(h)$ below which the witness certifies magic scales asymptotically as $T_W(h)\simeq 2h/\ln(4h/J)$ for $\Delta_{XZ}$, twice that for $\Delta_2$

}

\end{document}